\documentclass[12pt]{iopart}

\usepackage{graphicx}
\usepackage{bm}
\usepackage{amssymb} 
\usepackage{color}
\usepackage{hyperref}
\usepackage{ulem}

\newcommand{\nc}{\newcommand}
\nc{\rnc}{\renewcommand}
\nc{\nn}{\nonumber}

\nc{\g}{\gamma}
\nc{\om}{\omega}
\rnc{\b}{\beta}
\rnc{\th}{\theta}

\newcommand{\bra}{\langle}
\newcommand{\ket}{\rangle}
\nc{\vac}{|0\ket}
\nc{\vvac}{\bra0|}

\nc{\cd}{\cdots}
\nc{\sm}[2]{\sum_{#1=1}^{#2}}

\nc\hp{\hat{\psi}}
\nc\hpd{\hat{\psi}^\dagger}

\nc{\sg}{\sigma}
\nc{\lam}{\lambda}
\nc{\png}{\pngilon}
\rnc{\a}{\alpha}
\rnc{\b}{\beta}
\nc{\vp}{\varphi}
\nc{\kp}{\kappa}
\rnc{\i}{{\rm i}}
\rnc{\d}{{\rm d}}

    \newcommand{\Add}[1]{#1}	
    \newcommand{\Erase}[1]{\if0{#1}\fi}	

\begin{document}

\title[Dynamics of quantum double dark-solitons and  exact finite-size scaling of BEC 
]{Dynamics of quantum double dark-solitons and an exact finite-size scaling of Bose-Einstein condensation}

\author{Kayo Kinjo$^{1}$, Jun Sato$^{2}$ and Tetsuo Deguchi$^{3}$}

\address{$^{1}$ Department of Physics, Center for Soft Matter Physics, Ochanomizu University, Bunkyo-ku, Tokyo 112-8610, Japan\\
$^{2}$ Faculty of Engineering, Tokyo Polytechnic University, Atsugi 243-0297, Japan\\
$^{3}$ Department of Physics, Faculty of Core Research, Ochanomizu University, Bunkyo-ku, Tokyo 112-8610, Japan\\}
\ead{deguchi@phys.ocha.ac.jp}
\vspace{10pt}
\begin{indented}
\item[]June 2022 
\end{indented}

\begin{abstract}
We show several novel aspects in the exact non-equilibrium dynamics of quantum double dark-soliton states in the Lieb-Liniger model for the one-dimensional Bose gas with repulsive interactions. We also show an exact finite-size scaling of the fraction of the \Add{quasi-}Bose-Einstein condensation (BEC) in the ground state, which should characterize the quasi-BEC in quantum double dark-soliton states that we assume to occur in the weak coupling regime.  
First, we show the exact time evolution of the density profile in the quantum state associated with a quantum double dark-soliton by the Bethe ansatz. Secondly, we derive a kind of macroscopic quantum wave-function effectively by exactly evaluating the square amplitude and phase profiles of the matrix element of the field operator between the quantum double dark-soliton states. The profiles are close to those of dark-solitons particularly in the weak-coupling regime. 
Then, the scattering of two notches in the quantum double dark-soliton state is exactly demonstrated. 
It is suggested from the above observations that the quasi-BEC should play a significant role in the dynamics of quantum double dark-soliton states. 
If the condensate fraction is close to 1, the quantum state should be well approximated by the quasi-BEC state where the mean-field picture is valid. 
\end{abstract}

%
\vspace{2pc}

\submitto{\JPA}
%
%
%

\section{Introduction} 

The experimental realization of trapped atomic gases 
in one dimension (1D) has provided a new motivation for the study of strong correlations in fundamental quantum mechanical systems of interacting particles \cite{PhysRevLett.87.130402, PhysRevLett.87.160405, doi:10.1126/science.1100700, PhysRevLett.101.130401, becker2008oscillations}. 
Furthermore, the non-equilibrium dynamics of closed interacting quantum systems is now extensively studied in 1D by experiments and theories \cite{PhysRevLett.83.5198, katsimiga2017many,PhysRevLett.112.040402}.  
In many 1D quantum interacting systems 
quantum fluctuations may play a key role and often lead to subtle nontrivial effects.  
%
We thus expect that fundamental many-body properties such as the quasi-Bose-Einstein condensation (BEC) should play a key role in the nontrivial quantum dynamics such as quantum dark-solitons.  \Add{We shall define it shortly with the Penrose-Onsager criterion.}
 
Let us introduce a theoretical model for the 1D system of interacting bosons with repulsive short-range potentials. Here we call it the 1D Bose gas.  For simplicity we assume that the interactions are given by the delta-function potentials, since they give nontrivial effects in the 1D case although they are simple. For instance, the scattering length depends on the strength of the delta-function potential in 1D systems.  
We thus have the Lieb-Liniger model (LL model) as the system of the 1D Bose gas. 
The Hamiltonian of the LL model is given by \cite{PhysRev.130.1605, PhysRev.130.1616} 
\begin{eqnarray}
\mathcal{H}_{\rm{LL}} 
= - \sum_{j=1}^{N} {\frac {\partial^2} {\partial x_j^2}}
+ 2c \sum_{j < k}^{N} \delta(x_j-x_k) . 
\end{eqnarray}
Here \Add{$N$ denotes the number of bosons, and} we assume the periodic boundary conditions of the system size $L$ on the wave-functions. 
We employ a system of units with $2m=\hbar =1$, 
where $m$ denotes the mass of the particle. 
We recall that the coupling constant $c$ is positive. 
%
It is an exactly solvable model of the 1D quantum many-body system. It is known that all the eigenvectors are constructed by the Bethe-ansatz method \Add{\cite{dorlas1993orthogonality}}. Furthermore, the Gross-Pitaevskii (GP) equation appears as the Heisenberg equation of motion for the second-quantized Hamiltonian of the LL model. 
It is expressed in terms of the classical complex scalar field $\psi$  as follows \cite{pitaevskii2003bose}.  
\begin{eqnarray}
i \frac {\partial } {\partial t} \psi =- \frac {\partial ^2} {\partial x^2 } \psi + 2c |\psi|^2\psi -\mu\psi . \label{eq:GPE}  
\end{eqnarray}
We expect that the GP equation should play a central role in the long-distance mean-field behavior of the 1D Bose gas in some quantum state if the quasi-BEC occurs in the quantum state of the LL model especially in the weak-coupling regime. If it is the case, the solution of the GP equation should correspond to the macroscopic wave-function of the quasi-BEC state, and describe the quantum state well at least approximately. 

\Add{
We define the quasi-BEC by the criterion due to Penrose and Onsager 
\cite{leggett2006quantum, PhysRev.104.576} (see also Section \ref{subsec:criterion_BEC}).} 
\Add{Suppose that particle number $N$ is very large but finite. The density matrix at zero temperature is given by the ground state $|\lam\ket$  of the system 
as $\hat{\rho}=|\lam\ket\bra \lam |$. Then, we define the one-particle reduced density matrix by its partial trace with respect to all but one degree of freedom: $\hat{\rho}_1 = N\rm{tr}_{23\cd N}\hat{\rho}$. 
Let $N_0$ denote the largest eigenvalue of 
the one-particle reduced density matrix $\hat{\rho}_1$.  
If it is of order $N$, i.e., 
the ratio $n_0=N_0/N$ is nonzero and finite for large $N$, then we say that the system exhibits the quasi-BEC, and we call $n_0$ the condensate fraction. }

\Erase{Thus,} If the \Add{quasi-}BEC occurs in some quantum states of the LL model, we expect that the GP equation should play a central role for characterizing the quantum state, although it is only a partial differential equation for a complex scalar variable. 
In the present research, we assume that the quasi-BEC should occur if the coupling constant is small enough with respect to the system size or the number of bosons, and hence some solutions of the GP equation such as multiple dark-solitons can be compared with the density profiles of some quantum states in the quasi BEC of the 1D Bose gas. In fact, we shall show a finite-size scaling of the quasi BEC in the present research.

It should be \Add{emphasized} that such quantum states whose density profiles coincide with those of single dark-solitons of the GP equation have been constructed explicitly in the form of superposition of the yrast states in the Lieb-Liniger model \cite{Sato_2012}. 
The construction resolved a long standing problem suggested by Ishikawa and Takayama almost forty years ago \cite{doi:10.1143/JPSJ.49.1242(Ishikawa-Takayama)}. 
\Add{Here we remark that it was shown through the strong coupling limit} \Add{\cite{PhysRevLett.100.060401, PhysRevA.79.063616}}  
\Add{that the yrast states and the mean-field solitons are closely related to each other with respect to quantum numbers.} Furthermore, several significant properties in the non-equilibrium dynamics of a quantum single dark-soliton have been exactly investigated \cite{Sato_2016} and \Add{the generic and} the ideal Gaussian weights have been introduced \cite{PhysRevA.99.043632(Shamailov-Brand2019), kaminishi2020construction}. Moreover, \Add{the density and phase profiles of quantum states of double dark-solitons have been explicitly constructed \cite{kinjo2022quantum}, and} the phase shift has numerically been estimated in the scattering of two quantum dark-solitons  \cite{PhysRevResearch.4.L032047}.  

There is another aspect of quantum dark-soliton states.
Successive measurements of particle positions in the Lieb–Liniger model also leads to observing quantum dark-solitons numerically \cite{PhysRevA.92.032110(Syrwid-Sacha2015), Syrwid_2021}. There is a question of how the density profile of a superposition of yrast states is related to the successive measurements of particle positions. When the coupling constant $c$ is equal to zero it was analytically shown that the construction of the quantum dark-soliton state with the Gaussian weight \cite{kaminishi2020construction} is related to the particle position method \cite{PhysRevA.92.032110(Syrwid-Sacha2015)} \Add{as shown} in Ref. \cite{kaminishi2020construction}. When the coupling constant is small and nonzero: $c > 0$, an ansatz was proposed to bridge between the calculation of single-particle density and the particle position method \cite{PhysRevResearch.2.033368}.

In the present paper we show various novel aspects in the exact non-equilibrium dynamics of quantum double dark-solitons, which give pairs of notches in the density profiles, by explicitly constructing corresponding quantum states in the Lieb-Liniger model of the 1D Bose gas with the repulsive interactions.  
\Add{For instance, we exhibit the time evolution of the density profile of the double dark-soliton whose two notches are located at the same position, and that of the phase profiles of the quantum double dark-solitons. In particular, we give an example where the winding number of the phase profile changes during the scattering process of two notches. Furthermore, }
we also show an exact finite-size scaling of the fraction of the BEC 
\Add{for} the ground state. 
It should characterize the quasi-BEC which we assume to occur in quantum double dark-soliton states in the weak coupling regime.  
We show that if the coupling constant decreases as a power of the system size, condensate fraction does not vanish and remains constant when we send the system size to a very large value with fixed density. 
\Add{We recall that} if the condensate fraction is nonzero 
for a large particle number $N$, we call it \Add{the quasi-}BEC
\Add{by employing} the Penrose-Onsager criterion. 
It follows from it that the quasi-BEC occurs only 
if the coupling constant is very small with respect to 
the system size. Therefore quantum states of dark-solitons 
may appear particularly in the weak coupling regime.

Based on the definition of the quasi-BEC  
we derive a kind of macroscopic quantum wave-function 
by exactly deriving the amplitude and phase profiles 
of the matrix element of the bosonic field operator, by making use of Slavnov's formula of form factors \Add{\cite{slavnov1989calculation}}.
Here we recall that the bosonic field operator is defined in the second-quantized Hamiltonian of the Lieb-Liniger model \Add{\cite{korepin1993quantum}}. 


Let us briefly summarize the finite-size scaling of the quasi-BEC \Add{for the ground state, which we shall show in detail in Section 4.} The scaling behavior of the quasi-BEC in the 1D Bose gas is fundamental when 
we send particle number $N$ or system size $L$ to 
very large values. We define the interaction parameter $\gamma$ by $\gamma=c/n$ with coupling constant $c$ in the delta-function potentials and density $n=N/L$. We show that if $\gamma$ is given by 
a negative power of $N$, i.e. $\gamma=A/N^{\eta}$, 
condensate fraction $n_0$ is nonzero and constant for any large value of $L$ or $N$. We also show that exponent $\eta$ and amplitude $A$ are independent of density $n$, and evaluate them as functions of $n_0$. 
\Add{Thus, the } condensate fraction $n_0$ 
\Add{for the ground state} is given by 
a scaling function of variable $\gamma N^{\eta}$, \Add{which corresponds to amplitude $A$.}  
If the condensate fraction of a \Add{given} quantum state with large $N$ 
is nonzero in the 1D Bose gas, 
we suggest that the classical mean-field approximation 
such as the GP equation \Add{should be} valid 
for the state \cite{Sato_2012}.  
Furthermore, we show that the 1D Bose gas of 
a finite particle number may 
have the same condensate fraction for any large $L$ \Add{in the case of the ground state}. 

Finally, we mention some potentially relevant results in the following. 
For strong and intermediate interaction strengths, the Lieb-Liniger Gross-Pitaevski equation is introduced, which is an extension of the GP equation \cite{kopycinski2022beyond}. 
Associated with the quantum states of dark solitons, bound states of dark solitons are numerically studied by solving the GP equation \cite{PhysRevA.98.043612}, 
dynamics of a bright soliton in the quasi-BEC with time-dependent atomic scattering length in a repulsive parabolic potential \cite{liang2005dynamics},
quantized quasi-two-dimensional Bose-Einstein condensates with spatially modulated nonlinearity
\cite{wang2010quantized}, matter rogue wave in Bose-Einstein condensates with attractive atomic interaction \cite{wen2011matter}, exact soliton solutions, and nonlinear modulation instability in spinor Bose-Einstein condensates \cite{li2005exact}.

The contents of the paper consist of the following. 
In Section 2 we explain the Bethe ansatz and 
useful formulas for evaluating the form factors of 
the field operator. We also define the winding number for 
solutions of the GP equation under the periodic boundary conditions. 
In Section 3 we show the time evolution of the quantum double dark-soliton state constructed with equal weight for the following two cases: (i) The soliton positions $X_1$ and $X_2$ are different: $X_1=L/4$ and $X_2=3L/4$; (ii) the soliton positions are the same: $X_1=X_2=0$. We also show the time evolution of the quantum double dark-soliton state constructed with the Gaussian weights. Here, two notches have different speeds \Add{thanks to the Gaussian weights}, and we evaluate the phase shift in the collision of the two dark solitons.
\Add{We remark that two notches have mostly the same speed if the quantum double dark-soliton state is constructed with equal weight.}
In Section 4 we show the finite-size scaling behavior 
of the condensate fraction \Add{in the ground state for} 
the 1D Bose gas with repulsive interactions at zero temperature.
According to it, \Add{we can estimate that} 
the fraction of the quasi-BEC condensate \Erase{is} \Add{should be}  equal to 0.99 for the quantum double dark-soliton state with $N=L=20$ and $c=0.05$ studied in the present research.

\section{Method} 

\subsection{Bethe ansatz equations}
In the LL model, the Bethe ansatz offers an exact eigenstate 
with an exact energy eigenvalue 
for a given set of quasi-momenta 
$k_1, k_2, \ldots, k_N$ satisfying 
the Bethe ansatz equations (BAE) 
for $j=1, 2, \ldots, N$: 
\begin{eqnarray}
 k_j L = 2 \pi I_j - 2 \sum_{\ell \ne j}^{N} 
\arctan \left({\frac {k_j - k_{\ell}} c } \right) . 
\label{BAE} 
\end{eqnarray}
Here $I_j$'s are integers for odd $N$ and half-odd integers for even $N$. 
We call them the Bethe quantum numbers. 
The total momentum $P$ and the energy eigenvalue $E$ are expressed 
in terms of the quasi-momenta as 
\begin{eqnarray}
P=\sm{j}{N}k_j=\frac {2 \pi} L \sum_{j=1}^{N} I_j, \quad E=\sm{j}{N}k_j^2. 
\label{eq:momentum}
\end{eqnarray}
If we specify a set of Bethe quantum numbers 
$I_1<\cd<I_N$, the BAE in \Eref{BAE} have 
a unique real solution $k_1 < \cd < k_N$ \cite{korepin1993quantum, dorlas1993orthogonality}. 
In particular, the sequence of the Bethe quantum numbers 
of the ground state is given by 
\begin{eqnarray}
I_j=-(N+1)/2+j  \qquad \textrm{for $j \in \mathbb{Z}$ with $1 \leq j \leq N$}. 
\label{eq:bqn_}
\end{eqnarray}
The Bethe quantum numbers for low lying excitations are 
systematically derived by putting holes 
or particles in the 
perfectly regular ground-state sequence. 

\subsection{Coupling constant}

In the thermodynamic limit several physical quantities of the LL model are characterized by the single parameter $\gamma=c/n$, 
where $n=N/L$ is the density of particle number $N$.  
We often fix the particle-number density 
as $n=1$ throughout the present paper, and change 
coupling constant $c$ so that we have different values of $\gamma$.  

\subsection{Quantum double dark-soliton state} \label{subsec:qdoublesolstate}

\begin{figure}
    \centering
    \includegraphics[width=10cm]{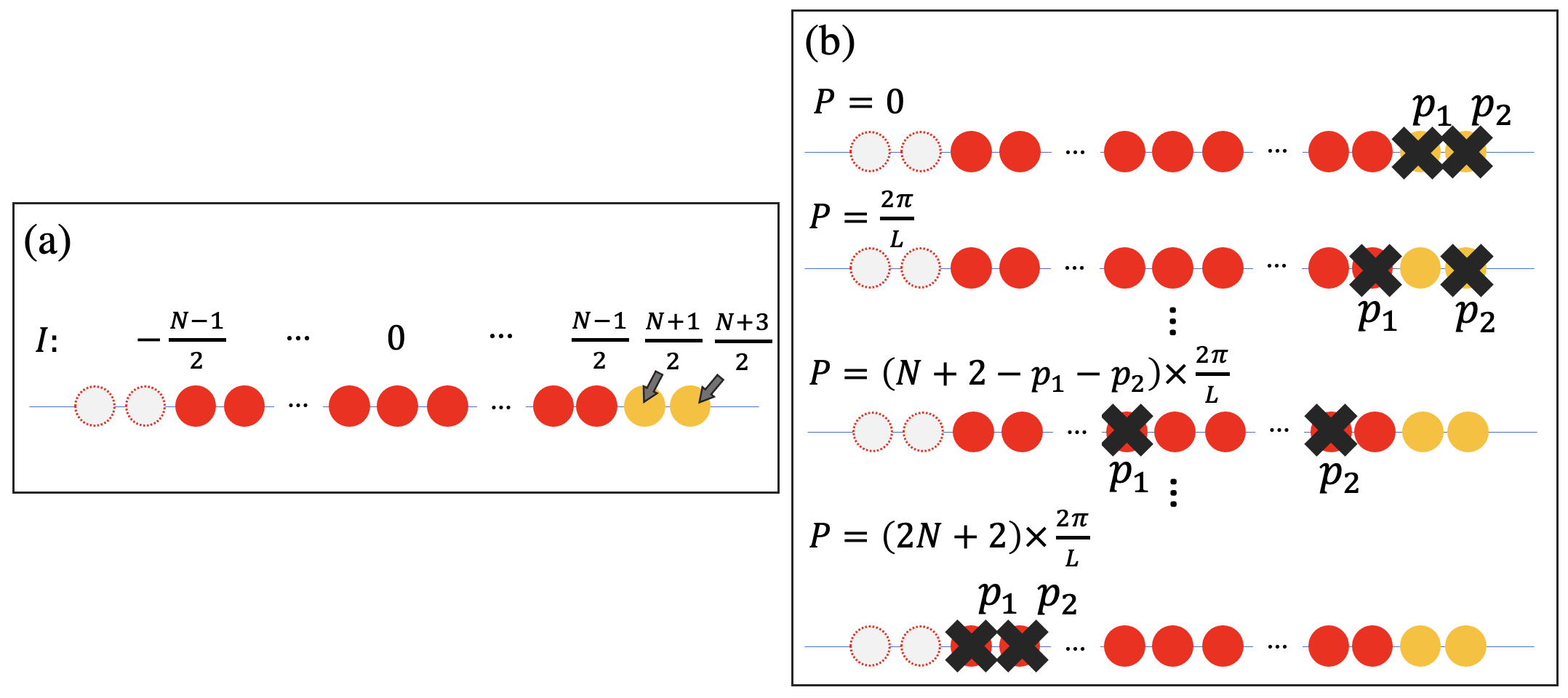}
    \caption{ (Color online) Configuration of Bethe quantum numbers for a quantum double dark-soliton state. (a) Configuration derived by adding two particles (yellow circles) to that of the ground state (red circles). (b) A series of configurations derived by punching two holes, which correspond to $\bm{p}=\{p_1,p_2\}$, in the configuration of the Bethe quantum numbers constructed in panel (a). \Add{The third configuration has two holes at $p_1$ and $p_2$.   }
    \Erase{Reproduced from \cite{kinjo2022quantum}.}}
    \label{fig:ConfigBetheQuantumNumber}
\end{figure}

A quantum state \Add{that has} \Erase{possessing the} two notches in both 
\Add{profiles of density} \Erase{its density profile} and square amplitude \Add{ of the matrix element of the field operator} was proposed in \cite{kinjo2022quantum}. \Add{We call it} \Erase{such  
quantum state is called} the quantum double dark-soliton state\Add{, and it is} given by the superposition of ``two-hole" excitation states 
\Add{as follows.}
\begin{eqnarray}
|X_1, X_2 , N\ket := \frac{1}{\sqrt{\mathcal{M}_N}}\sum_{\bm{p}\in \bm{P}_N}e^{i(p_1X_1+p_2X_2)}|p_1,p_2,N\ket
\label{eq:XN}
\end{eqnarray}
with a normalization factor $\mathcal{M_N}$ \Add{for $N$ particles}. 
\Erase{Here $|p_1,p_2,N\ket$ denotes a two-hole excitation for which the configuration of the Bethe quantum numbers are determined by the two holes $p_1$ and $p_2$, and $\bm{P}$ denotes the set of all allowed pairs of two holes for $N$ particles. \Fref{fig:ConfigBetheQuantumNumber} illustrates how to punch holes in the series of the Bethe quantum numbers.}
\Add{The quantum state $|p_1,p_2,N\ket$ is characterized by a configuration of Bethe quantum numbers that has two vacancies located at $p_1$ and $p_2$ in the series of the Bethe quantum numbers,  which is  illustrated in \Fref{fig:ConfigBetheQuantumNumber} (a). This configuration represents the Bethe quantum numbers of the ground state of $N$ particles along with those of additional two particles.
 In Equation (\ref{eq:XN}), the two holes are denoted as $\bm{p}:=\{p_1, p_2\}$ hereinafter, and the set of all allowed pairs of two holes is represented by $\bm{P}_N$. For example, when the number of particles $N=5$, $\bm{P}_N = \left\lbrace\{-2,-1\},\{-2,0\},\{-2,1\},\{-2,2\},\cdots, \{2,4\}, \{3,4\}\right\rbrace$, which has $|\bm{P}_N|=21$ elements. In the panel (b) of \Fref{fig:ConfigBetheQuantumNumber} some configurations with two holes $p_1$ and $p_2$ are exhibited. In the third configuration, two holes $p_1$ and $p_2$ are located in its middle part of the series which corresponds to the ground state of $N$ particles. 
}
\Add{Here we remark that in order for two notches have  positive velocities we derive two hole excitations derived from the configuration constructed by adding two particles to the right of the "Fermi momentum" as shown in \Fref{fig:ConfigBetheQuantumNumber} (a). If we add the two particles to the right 
and left of the "Fermi momentum" symmetrically, then the sum of the momenta vanishes. }

The density profile of this state 
$\bra X_1, X_2, N|
\hat{ \psi}^{\dagger}(x) \hat{ \psi}(x)
|X_1, X_2 ,N\ket$ 
shows the two density notches at the positions $x=X_1,X_2$, 
which coincides with the squared amplitude of the elliptic soliton \cite{kinjo2022quantum}. 
Here, by the determinant formula 
for the norms of Bethe eigenstates \cite{gaudin1983fonction, Korepin1982} 
we can effectively evaluate the matrix element
\begin{eqnarray}
&\bra X_1, X_2, N|
\hat{ \psi}^{\dagger}(x) \hat{ \psi}(x)
|X_1, X_2, N\ket
=
\frac1{\mathcal{M}_N}
\sum_{\Add{\bm{p},\bm{p'} \in \bm{P}_N}}
e^{i(P-P')x}e^ {i(p_1X_1+p_2X_2)}
\nonumber\\&
\times e^ {-i(p'_1X_1+p'_2X_2)}
\bra p'_1,p'_2, N| \hat{\psi}^\dagger(0)\hat{\psi}(0)| p_1,p_2, N\ket.
\label{eq:DensityProfile}
\end{eqnarray}
\Add{Here, $P$ and $P'$ in an exponential term denote the total momentum of the state $|p_1,p_2,N\ket$ and $|p'_1,p'_2,N\ket$ calculated through Equation (\ref{eq:momentum}), respectively. The sum in the above equation is taken over all pairs of $\bm{p} = \{p_1, p_2\}$ and $\bm{p'} = \{p'_1, p'_2\}$ that belong to the set $\bm{P}_N$.}
\Erase{and that of} \Add{The matrix element of} the form factors of the density operator \cite{slavnov1989calculation, Caux_2007,kojima1997determinant} \Add{is given by}
\begin{eqnarray}
&\bra p'_1,p'_2, N| \hat{\psi}^\dagger(0)\hat{\psi}(0)| p_1,p_2, N\ket
\nonumber\\&
=(-1)^{N(N+1)/2}(P-P')
\left(\prod^N_{j,\ell=1}\frac{1}{k'_j-k_\ell}\right)
\left( \prod^N_{j>\ell}k_{j,\ell}k'_{j,\ell}\sqrt{\frac{{\hat K}(k'_{j,\ell})}{{\hat K}(k_{j,\ell})} } \right)
\nonumber\\& \times
\frac{\det U(k,k')}{\sqrt{\det G(k)\det G(k')}}, 
\label{eq:Slavnov_density}
\end{eqnarray}
where the quasimomenta $\{k_1,\cdots,k_N\}$ and $\{k'_1,\cdots,k'_N\}$ 
give the eigenstates \Erase{$|p\ket$ and $|P'\ket$} \Add{$|p_1, p_2, N\ket$ and $|p'_1, p'_2, N\ket$}, respectively. 
We use the abbreviations $k_{j,\ell}:=k_j-k_\ell$ and $k'_{j,\ell}:=k'_j-k'_\ell$. 
The kernel ${\hat K}(k)$ is defined by ${\hat K}(k)=2c/(k^2+c^2)$. 
The matrix $G(k)$ is called the Gaudin matrix, whose $(j,\ell)$ th element is given by 
\begin{equation} 
G(k)_{j,\ell}=\delta_{j,\ell}
\left[L+\sum_{m=1}^N {\hat K}(k_{j,m})\right]-{\hat K}(k_{j,\ell})
\quad \mbox{for} \quad j, \ell=1, 2, \cdots, N . \label{eq:Gaudin}
\end{equation} 
The matrix elements of the $(N-1)$ by $(N-1)$ matrix $U(k,k')$ are given by 
\begin{eqnarray}
U(k,k')_{j,\ell}
&=2\delta_{j\ell}\textrm{Im}
\left[\prod^N_{a=1}
\frac{k'_a-k_j + i c}{k_a-k_j + i c}
\right]
\nonumber\\&
+\frac{\prod^N_{a=1}(k'_a-k_j)}{\prod^N_{a\neq j}(k_a-k_j)} 
\left({\hat K}(k_{j,\ell})-{\hat K}(k_{N,\ell})\right). 
\label{eq:matrixU_density}
\end{eqnarray}

We have also considered the matrix element of the single field operator
\begin{eqnarray}
\nn&
\bra X_1, X_2, N-1| \hat{\psi}(x)| X_1, X_2, N\ket\\ \nn
&=\frac{1}{\sqrt{\mathcal{M}_N\mathcal{M}_{N-1}}}
\sum_{\Add{\bm{p'}\in \bm{P}_{N-1}}}\sum_{\Add{\bm{p}\in \bm{P}_{N}}}
e^{i(P-P')x}e^ {i(p_1X_1+p_2X_2)}e^ {-i(p'_1X_1+p'_2X_2)}\\
&\times
\bra p'_1, p'_2, N-1| \hat{\psi}(0) |p_1, p_2, N\ket ,
\label{eq:matrix}
\end{eqnarray}
where $P$\Erase{$=2\pi p/L$} and $P'$\Erase{$=2\pi p'/L$} denote the total momenta 
of the state \Erase{normalized Bethe eigenstates in the type II branch 
$|P, N\ket$ and $|P', N\ket$} \Add{$|p_1, p_2, N\ket$ and $|p'_1, p'_2, N-1\ket$}, respectively. 
The determinant formula is given by
\cite{gaudin1983fonction, Korepin1982, slavnov1989calculation, slavnov1990, Caux_2007, kojima1997determinant}
\begin{eqnarray}
\nn
&\bra p'_1,p'_2, N| \hat{\psi}(0)| p_1,p_2, N\ket
=(-1)^{N(N+1)/2+1}
\left(
\prod^{N-1}_{j=1}
\prod^N_{\ell=1}
\frac{1}{k'_j-k_\ell}\right) 
\nn\\
\times
&\left( 
\prod^N_{j>\ell}k_{j,\ell}\sqrt{k_{j,\ell}^2+c^2} 
\right)
\left( 
\prod^{N-1}_{j>\ell}\frac{k'_{j,\ell}}{\sqrt{(k'_{j,\ell})^2+c^2}} 
\right)
\frac{\det \widehat{U}(k,k')}{\sqrt{\det G(k)\det G(k')}}, 
\label{eq:Slavnov_field}
\end{eqnarray}
where the quasi-momenta $\{k_1,\cdots,k_N\}$ and $\{k'_1,\cdots,k'_{N-1}\}$ 
give the eigenstates \Erase{$|P, N\ket$ and $|P', N-1\ket$} \Add{$|p_1, p_2, N\ket$ and $|p'_1, p'_2, N-1\ket$}, respectively. 
\Erase{Here we remark that the symbol $X^{'}$ corresponds to an integer $q^{'}$.}
We recall that the matrix $G(k)$ denotes the Gaudin matrix, whose $(j,\ell)$th element is 
given in Equation \eref{eq:Gaudin}. 
%
%
The matrix elements of the $(N-1)$ by $(N-1)$ matrix $\widehat{U}(k,k')$ are given by 
\begin{eqnarray}
\widehat{U}(k,k')_{j,\ell}&=2\delta_{j\ell}\textrm{Im}
\left[
\frac{\prod^{N-1}_{a=1}(k'_a-k_j + ic)}{\prod^N_{a=1}(k_a-k_j + ic)}
\right]
 \nn\\&
+\frac{\prod^{N-1}_{a=1}(k'_a-k_j)}{\prod^N_{a\neq j}(k_a-k_j)}
\left({\hat K}(k_{j,\ell})-{\hat K}(k_{N,\ell})\right). 
\label{eq:matrixU_field}
\end{eqnarray}

\subsection{One-particle reduced density matrix}
The matrix element of the one-particle reduced density matrix, $\rho_1(x,y):=\bra x|\hat{\rho}_1|y\ket$, 
 for a quantum system is expressed 
as a correlation function in the ground state 
$| \lambda \rangle$: 
\begin{eqnarray}
\rho_1(x,y) = \bra \lam |\hat{\psi}^\dagger(y)\hat{\psi}(x)| \lam \ket. 
\end{eqnarray} 

In the LL model we can numerically 
evaluate the correlation function 
by the form factor expansion. 
Inserting the complete system of eigenstates, 
$\sum_{\mu} |\mu \rangle \langle \mu|$, we have 
\begin{eqnarray}
\rho_1(x,y)
=\sum_\mu
e^{\i(P_\mu-P_\lam)(y-x)}
| \bra \mu |\hat{\psi}(0)| \lam \ket |^2, 
\label{eq:sum-ff}
\end{eqnarray}
where $P_\mu$ denotes the momentum eigenvalues of eigenstates $|\mu\ket$. 
Each form factor in the sum (\ref{eq:sum-ff}) is expressed as a  
product of determinants by making use of the determinant formula 
for the norms of Bethe eigenstates \cite{gaudin1983fonction} and 
that for the form factors of the field operator 
\cite{slavnov1989calculation, kojima1997determinant, Caux_2007}: 
\begin{eqnarray}
\nn
&\bra\mu|\hp(0)|\lam\ket
=(-1)^{N(N+1)/2+1}
\left(
\prod^{N-1}_{j=1}
\prod^N_{\ell=1}
\frac{1}{k'_j-k_\ell}\right) 
\left( 
\prod^N_{j>\ell}k_{j,\ell}\sqrt{k_{j,\ell}^2+c^2} 
\right)\\
&\times
\left( 
\prod^{N-1}_{j>\ell}\frac{k'_{j,\ell}}{\sqrt{(k'_{j,\ell})^2+c^2}} 
\right)
\frac{\det U(k,k')}{\sqrt{\det G(k)\det G(k')}} \, , 
\label{eq:Slavnov}
\end{eqnarray}
where the quasi-momenta $\{k_1,\cd,k_N\}$ and $\{k'_1,\cd,k'_{N-1}\}$ 
give the eigenstates $|\lam\ket$ and $|\mu\ket$, respectively. 
Here we have employed the abbreviated symbols 
$k_{j,\ell}:=k_j-k_\ell$ and $k'_{j,\ell}:=k'_j-k'_\ell$. 
The matrix $G(k)$ is the Gaudin matrix, whose $(j,\ell)$th element is given by
$G(k)_{j,\ell}=\delta_{j,\ell}\left[L+\sum_{m=1}^NK(k_{j,m})\right]-K(k_{j,\ell})$
for $j, \ell=1,2,\cd,N$, where the kernel $K(k)$ 
is defined by $K(k)=2c/(k^2+c^2)$. 
The matrix elements of the $(N-1)$ by $(N-1)$ matrix $U(k,k')$ are given by 
\cite{Caux_2007, slavnov1989calculation, kojima1997determinant, gaudin1983fonction} 
\begin{eqnarray}
\nn
U(k,k')_{j,\ell}
&=2\delta_{j\ell}\textrm{Im}
\left[
\frac{\prod^{N-1}_{a=1}(k'_a-k_j + ic)}{\prod^N_{a=1}(k_a-k_j + ic)}
\right]
\\
&+\frac{\prod^{N-1}_{a=1}(k'_a-k_j)}{\prod^N_{a\neq j}(k_a-k_j)}
\left(K(k_{j,\ell})-K(k_{N,\ell})\right) . 
\label{eq:matrixU}
\end{eqnarray}

\Add{For the ground state $| \lambda \rangle$  
we have shown that the sum of the form factor expansion is almost saturated for the one-particle and one-hole (1p1h) excitations together with  two-particles and two-holes (2p2h) excitations. The saturation rate is explicitly presented in Table \ref{tab:saturation_rate} of Section \ref{subsec:saturation_rate}.  However, for excited states the saturation rate has not been evaluated. It should be technically nontrivial to evaluate it for excited states. 
For the quantum states of double dark-solitons,  
we suggest that the saturation rate should be close to one in the weak coupling case in the form factor expansion up to some excitations with relatively small numbers of particles and holes. It is based on the observation that the density profiles of quantum double dark-soliton states are similar to those of the double dark-solitons of the GP equation, as we shall show in Section 3. } 

\subsection{Winding number}
We introduce the winding number $J$ associated with solutions of the GP equation under the periodic boundary conditions. Let us assume that a solution of the GP equation $\phi(x) = \sqrt{\rho(z)}\exp[i\varphi(x)]$ satisfies the periodic boundary conditions:
\begin{eqnarray}
    \varphi(x+L) = \varphi(x) + 2\pi J
    \label{eq:pbc_winding}
\end{eqnarray}
where $J$ is an arbitrary integer. The integer $J$ is called the winding number \cite{PhysRevLett.100.060401, PhysRevA.79.063616}. In the previous study, we constructed the quantum single dark-soliton with a nonzero-winding number.

\section{Dynamics of quantum double dark-soliton} 


\subsection{Time evolution of quantum double dark-soliton state constructed with equal weight}\label{subsec:timeevolve_qsol_wEqualweight}
By \Add{making use of} \Erase{applying} 
the time dependent field operator $\hat{\psi}(x,t)$, 
the \Add{local} density \Erase{profile} and the matrix element of the quantum state 
\Erase{evolve in time} \Add{at a given time $t$ are expressed} as follows.
\begin{eqnarray}
    \nn
    \rho_Q(x,t)
    &:=\langle X_1, X_2, N|\hat{\rho}(x, t)| X_1, X_2, N\rangle\\
    \nn
    &=\frac{1}{\mathcal{M}_N}\sum_{\bm{p},\bm{p'}\in \bm{P}_N}
     e^{i(P-P')x}e^ {i(p_1X_1+p_2X_2)}e^ {-i(p'_1X_1+p'_2X_2)}
    e^{ \left[-\mathrm{i}\left(E-E^{\prime}\right) t\right]}\\
    &\times\bra p_1',p_2',N|\hat{\rho}(0,0)|p_1,p_2,N\ket,
    \label{eq:time_den}
\end{eqnarray}
\begin{eqnarray}
    \nn
    \psi_Q(x,t) &:= \bra X_1,X_2,N-1|\hat{\psi}(x,t)|X_1,X_2,N\ket\\
    \nn
	&=\frac{1}{\sqrt{\mathcal{M}_{N-1}\mathcal{M}_{N}}}\sum_{\bm{p'}\in \bm{P}_{N-1}}\sum_{\bm{p}\in \bm{P}_{N}} e^{i(P-P')x}e^ {i(p_1X_1+p_2X_2)}e^ {-i(p'_1X_1+p'_2X_2)}
	\\
	&\times e^{-i(E-E')t}\bra p_1',p_2',N-1|\hat{\psi}(0,0)|p_1,p_2,N\ket,
	\label{eq:time_matele}
\end{eqnarray}
where $E$ is the energy of the state $|p_1,p_2,N\ket$, and $\hat{\rho}(x,t) =\hat{\psi}^\dagger(x,t)\hat{\psi}(x,t)$ \Add{denotes the local density operator}. We have obtained the exact expressions of the time evolution in Equations \eref{eq:time_den} and \eref{eq:time_matele} since the Bethe ansatz method gives the exact energies for the quantum state $|X_1,X_2,N\ket$. 
 
\subsubsection{Quantum dark-soliton \Erase{positions}
located at $X_1=L/4$ and $X_2=3L/4$ initially} \label{subsubsec:uniform_samedistance}

\begin{figure}[h]
    \centering
    \includegraphics[width=0.83\linewidth]{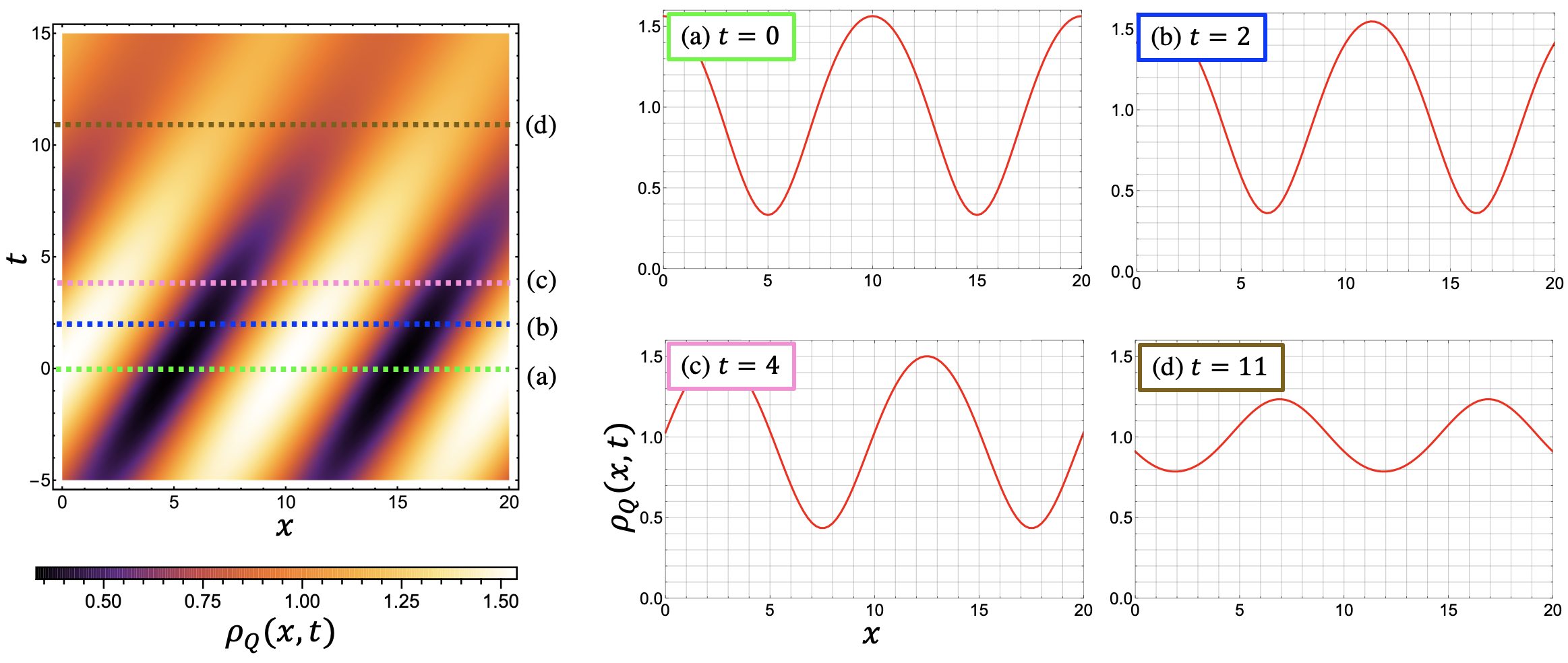}
    \caption{ (Color online)  \Add{2D density plot of local density $\rho_Q(x,t)$ in the space-time diagram} 
   \Erase{Time evolution of the density profile $\rho_Q(x,t)$} 
   (left panel) and snapshots \Add{of the density profile 
   of $\rho_Q(x,t)$} at $t=0,\ 2,\ 4$, and 11, in panels (a), (b), (c), and (d), respectively \Add{(right panels)}, for the quantum double dark-soliton state  \Erase{$X_1=\frac{L}{4}, X_2=\frac{3L}{4},$} with $N=L=20$  
    and $c=0.05$. \Add{The corresponding quantum state is given by Equation (\ref{eq:XN}) with $X_1=\frac{L}{4}$ and $X_2=\frac{3L}{4}$}. In the left panel, the vertical \Add{(horizontal)}
axis corresponds to time \Add{$t$ (coordinate $x$).} \Erase{, and the horizontal one to the $x$ coordinate.} Panel (a) shows the density profile at $t=0$, which corresponds to the green dotted line in \Add{the 2D density plot} of the left panel.  
\Erase{the density plot.} Panels (b), (c), and (d) show the density profile at $t=2, 4$, and 11, which correspond to the blue, pink, and ocher dotted lines in the \Add{2D density plot} (left panel) \Erase{of the density plot}, respectively. The frames of the legends of panels (a), (b), (c), and (d) correspond to the green, blue, pink, and ocher colors, respectively.
}
    \label{fig:dynamics_N20_c005_den}
\end{figure}

\Fref{fig:dynamics_N20_c005_den} shows the time evolution of the density profile \Add{, i.e., the graph of $\rho_Q(x,t)$ versus $x$ at a given time $t$, }   \Add{for the quantum double dark-soliton state} with \Add{initial} soliton positions $X_1=\frac{L}{4}$ and $X_2=\frac{3L}{4}$ \Erase{at $t=0$} under the periodic boundary conditions. 
\Add{We call the plot 
in the left panel of \Fref{fig:dynamics_N20_c005_den} 
the two-dimensional (2D) density plot of the local density. 
Here, the value of the local density $\rho_Q(x, t)$ at position $x$ and time $t$ is expressed by the brightness of the point at $(x,t)$ in the space-time diagram,  where the horizontal axis corresponds to the $x$ coordinate, while the vertical axis to time $t$. 
In the right panels of \Fref{fig:dynamics_N20_c005_den} 
snapshots of the density profile of $\rho_Q(x,t)$ 
at $t=0, 2, 4$, and 11 are plotted. 
}

We note that the density profile shown in panel (a) of \Fref{fig:dynamics_N20_c005_den} is identical to the upper-left panel of Figure 9 for $c=0.05$ in Ref. \cite{kinjo2022quantum}. 
\Add{In the latter panel it was shown that the density profile of the quantum double dark-soliton state completely coincides with the density profile of the elliptic double dark-soliton solution of the GP equation.} 
Thus, at $t=0$, the density profile of the quantum double dark-soliton state coincides with that of the elliptic soliton solution of the GP equation. 

The positions of notches are \Add{expressed by} \Erase{are plotted in} 
the areas of the darker color \Add{in the 2D density plot} at the left panel of \Fref{fig:dynamics_N20_c005_den}. 
\Add{
The trajectories of the positions of the two notches in the density profile are given by two parallel linearly elongated regions in the diagram of time $t$ and coordinate $x$, as shown in the left panel of \Fref{fig:dynamics_N20_c005_den}. Thus, the two notches moves at the same velocity in the positive $x$ direction.  In the snapshots of the density profiles, the soliton notches are gradually filled, i.e., they become shallower in time evolution, as shown in panels (a), (b), (c), and (d) of \Fref{fig:dynamics_N20_c005_den}.}
That is, the distance between the bottoms of the notches is kept constant through the time evolution, while the depths of the notches become smaller. \Add{Here we have defined the depth of a notch by the difference between the largest and smallest values in the density profile.} For example, at $t=11$, the notches are located at $x_1=1.9115$ and $x_2=11.9115$, and the distance between the two notches is given by $\Delta x = x_1 - x_2=10=L/2$, which is \Add{equal to} that of $t=0$.

\Add{
It was reported in Ref. \cite{Syrwid2016} that quantum double dark-solitons with notches of almost the same depths can appear again after their depths of notches become much smaller over a time scale of $1/c$. However, the quantum double dark-soliton states constructed in the present research do not show this reappearing or recurrent behavior in time evolution. Once the soliton notches in the density profile are completely filled, i.e., their depths vanish, the density profile remains flat and uniform in time evolution, as illustrated in Figure \ref{fig:dynamics_N20_c005_den}. We note that the construction of the quantum soliton in Ref. \cite{Syrwid2016} is different from that of the present research, and also that the number of particles is equal to $N=8$ in Ref. \cite{Syrwid2016}, which is smaller than $N=20$ for the system in Figure \ref{fig:dynamics_N20_c005_den}.}

\begin{figure}[h]
    \centering
    \includegraphics[width=0.83\linewidth]{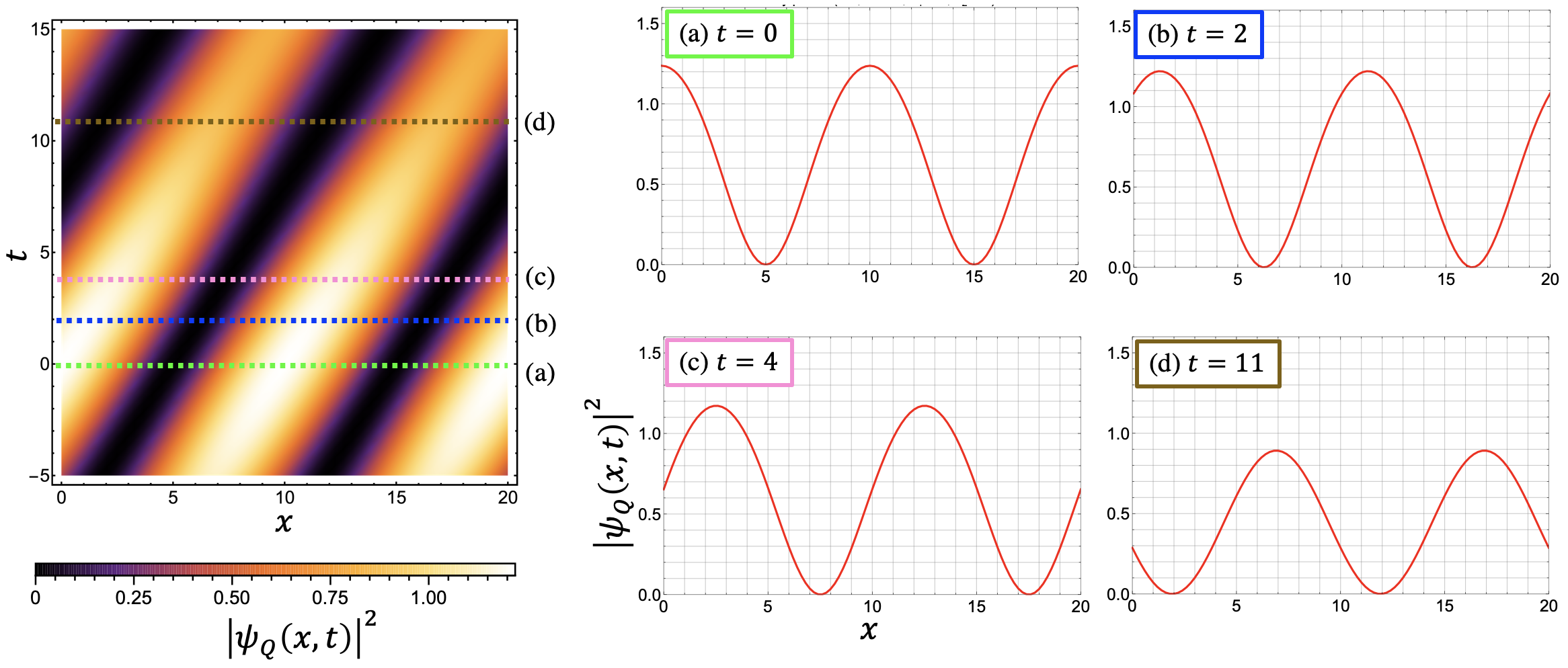}
    \caption{ (Color online)  \Add{2D density plot of} \Erase{Time evolution of} the square amplitude $|\psi_Q(x,t)|^2$ of 
    the matrix element $\psi_Q(x,t)$ \Add{in the space-time diagram} (left panel) 
    and \Erase{its} snapshots 
    \Add{of the square amplitude profile (i.e., $|\psi_Q(x,t)|^2$ versus $x$)} at $t=0, 2, 4$ and 11,  
     in panels (a), (b), (c), and (d), respectively, \Add{(right panels),} for \Erase{$X_1=\frac{L}{4}, X_2=\frac{3L}{4},$} $N=L=20$ and $c=0.05$.
    \Erase{(right panels)} 
    \Add{The corresponding quantum states are given by Equation (\ref{eq:XN}) with $X_1=\frac{L}{4}$ and $X_2=\frac{3L}{4}$ 
    for $N-1$ and $N$ particles}.
    \Add{In the left panel, the vertical (horizontal)
axis corresponds to time $t$ (coordinate $x$).} 
Panels (a), (b), (c), and (d) exhibit the square amplitude profile at $t=0, 2, 4$, and 11, which correspond to the green, blue, pink, and ocher dotted lines in the \Add{2D density plot} (left panel).  
    }
 \label{fig:dynamics_N20_c005_field}
\end{figure}

\Add{The notches in the density profile of $\rho_Q(x,t)$
and \Add{those in the profile of} the square amplitude  $|\psi_Q(x,t)|^2 $ of matrix element $\psi_Q(x,t)$  
exhibit different decaying behaviors in time evolution.} 
Figure \ref{fig:dynamics_N20_c005_field} shows the \Add{time} evolution of the square amplitude \Add{profile} of matrix element $\psi_Q(x,t)$ with \Add{initial} soliton positions $X_1=\frac{L}{4}$ and $X_2=\frac{3L}{4}$ \Erase{at $t=0$} under the periodic boundary conditions. The average density is decreasing in the time evolution of the \Add{profile of the} square amplitude $|\psi_Q(x,t)|^2 $ in \Fref{fig:dynamics_N20_c005_field}, while the notches in \Fref{fig:dynamics_N20_c005_den} are \Erase{being} filled gradually. 
\Add{In the density profile, the average density is kept constant} as time $t$ increases\Add{, since the density is conserved as a whole for any time $t$: $\int_0^L dx \rho_Q(x,t) = N$}. \Add{On the other hand,} we suggest that \Add{the amplitude of the matrix element between the two different quantum states of double dark-soliton should gradually decrease and finally vanish in time evolution, 
since they have different energies and particle numbers. }
\Erase{the main components of which the quantum soliton state consists should decrease in the case of the square amplitude, whereas the density is conserved as a whole for any $t$: $\int_0^L dx \rho_Q(x,t) = N $.}

\Erase{In the same way as the density profile,}
 \Add{In the 2D density plot at the left panel of Figure  \ref{fig:dynamics_N20_c005_field},  
the trajectories of notches \Add{in the space-time diagram} 
are \Erase{shown with} \Add{depicted by linearly elongated parallel regions} \Erase{lines} with darker color}
\Erase{in the left panel of Figure \ref{fig:dynamics_N20_c005_field}}. \Erase{We observed that} The values at the bottoms of the notches are almost equal to zero constantly \Add{in time evolution} in panels (a), (b), (c), and (d) of \Fref{fig:dynamics_N20_c005_field}. Consequently, \Fref{fig:dynamics_N20_c005_field} shows \Add{the trajectories of} the notches more clearly than \Fref{fig:dynamics_N20_c005_den}, \Add{as depicted in the 2D density plot at the left panel}.

\begin{figure}[h]
    \centering
    \includegraphics[width=0.6\linewidth]{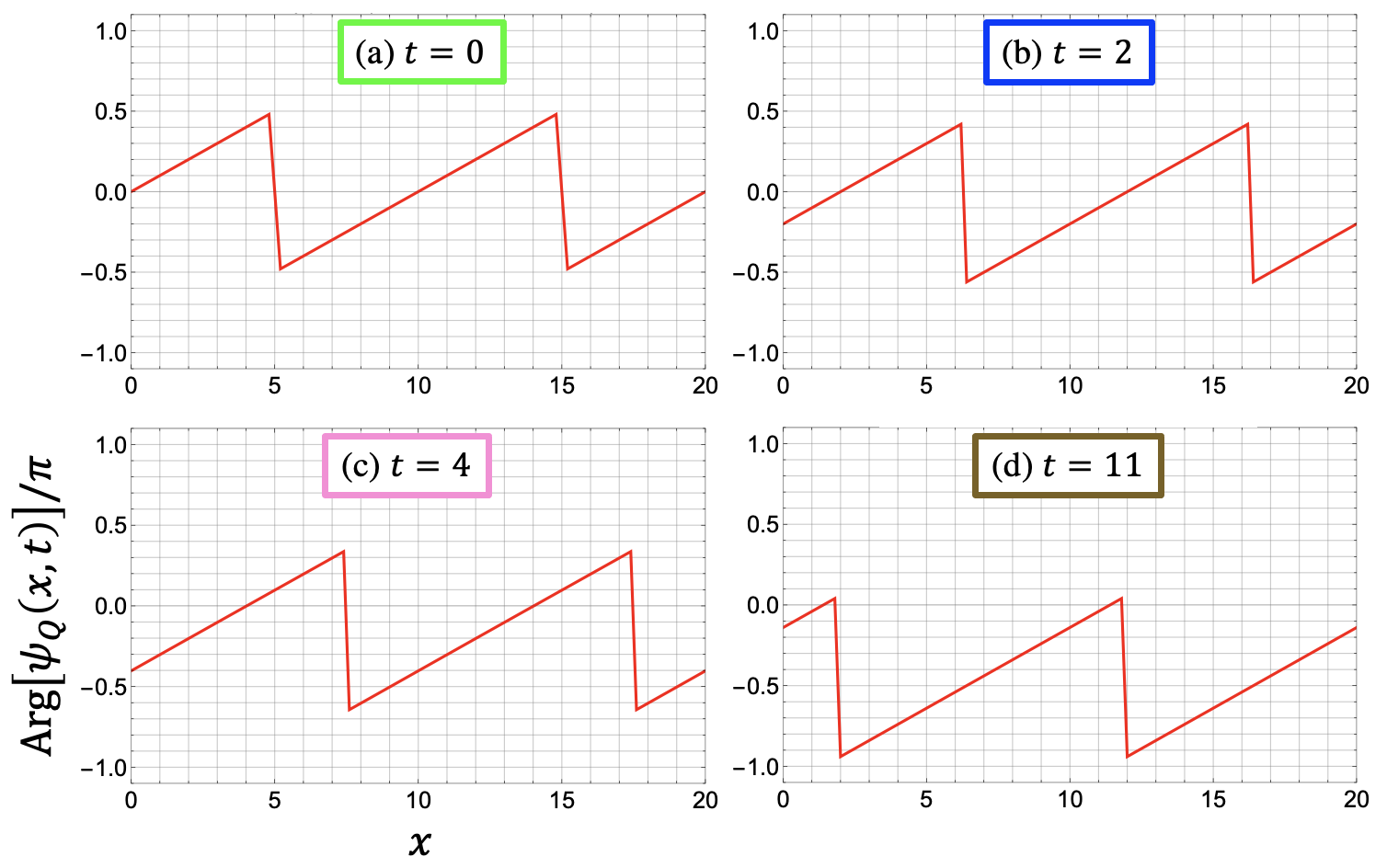}
    \caption{ (Color online) Snapshots of \Erase{the time evolution of} the phase profile for the matrix elements $\psi_Q(x,t)$ at $t=0, 2, 4$, \Add{and} 11, for $N=L=20$ \Erase{$X_1=\frac{L}{4}, X_2=\frac{3L}{4}$,} and $c=0.05$. \Add{The corresponding quantum states are given by Equation (\ref{eq:XN}) with $X_1=\frac{L}{4}$ and $X_2=\frac{3L}{4}$ for $N-1$ and $N$ particles.}}
    \label{fig:N20_c005_uni_Phase_time}
\end{figure}

\Erase{The time evolution of} \Add{The snapshots of} the phase profile \Add{at different times in time evolution} are shown in \Fref{fig:N20_c005_uni_Phase_time}. Here we remark that the phase is given by the argument of the matrix element of \Eref{eq:time_matele} \Add{as a complex number}. In \Fref{fig:N20_c005_uni_Phase_time} the abrupt jumps \Add{of the phase profile} are located at the positions of the notches in \Fref{fig:dynamics_N20_c005_field}. The abrupt jumps \Add{of the phase profile} move with the same constant velocity as \Add{the notches in} the square amplitude profile. \Add{Furthermore,} the whole phase profile is \Add{gradually} shifted toward the negative direction in time evolution. Moreover, \Erase{we observed that} the \Erase{whole} shape \Add{of the phase profile as a whole} \Erase{retains} \Add{remains the same} at least up to $t=40$.

At \Add{the initial time} $t=0$, \Add{the profiles of} the square amplitude and the phase of the matrix element $\psi_Q(x,t)$ shown in panels (a) of \Fref{fig:dynamics_N20_c005_field} and \Fref{fig:N20_c005_uni_Phase_time} are identical to those of Figures 10 and 11 for $c=0.05$ in Ref. \cite{kinjo2022quantum}, respectively. \Add{Panel (a) of \Fref{fig:dynamics_N20_c005_field}, the square amplitude profile of the matrix element, corresponds to the panel of $c=0.05$ in Figure 10 of Ref. \cite{kinjo2022quantum}, where it was shown that the square amplitude profile of the classical and quantum double dark-soliton overlap completely. Panel (a) of \Fref{fig:N20_c005_uni_Phase_time}, the phase profile of the matrix element, corresponds to the panel of $c=0.05$ in Figure 11 of Ref. \cite{kinjo2022quantum}, where the phase profiles of the classical and quantum double dark-solitons overlap completely.}

However, \Add{
the time evolution of the phase profile in the quantum double dark-soliton state is different from that of the elliptic dark-soliton solution, which is given by the travelling wave solution of the GP equation. }
\Erase{it follows from the phase field of}
\Add{We recall that the phase profile in the quantum double dark-soliton is gradually shifted toward the negative direction in time evolution 
in \Fref{fig:N20_c005_uni_Phase_time}, while} 
the phase profile of the travelling wave solution is not \Add{shifted.}
Thus, the time evolution of the quantum dark-solitons that we have constructed is slightly different from that of the 
\Add{classical} elliptic soliton solution.

\Add{We remark that two notches have mostly the same velocity as shown in Figures \ref{fig:dynamics_N20_c005_den} and \ref{fig:dynamics_N20_c005_field} for the quantum double dark-soliton constructed with equal weight. In Section \ref{subsec:qsol_g} we shall show that two notches have different \Erase{speeds} \Add{velocities} 
for the quantum double dark-soliton state constructed with the Gaussian weights.}

\subsubsection{Quantum dark-soliton positions located at $X_1=X_2=0$ initially}

\begin{figure}[ht]
    \centering
    \includegraphics[width=0.83\linewidth]{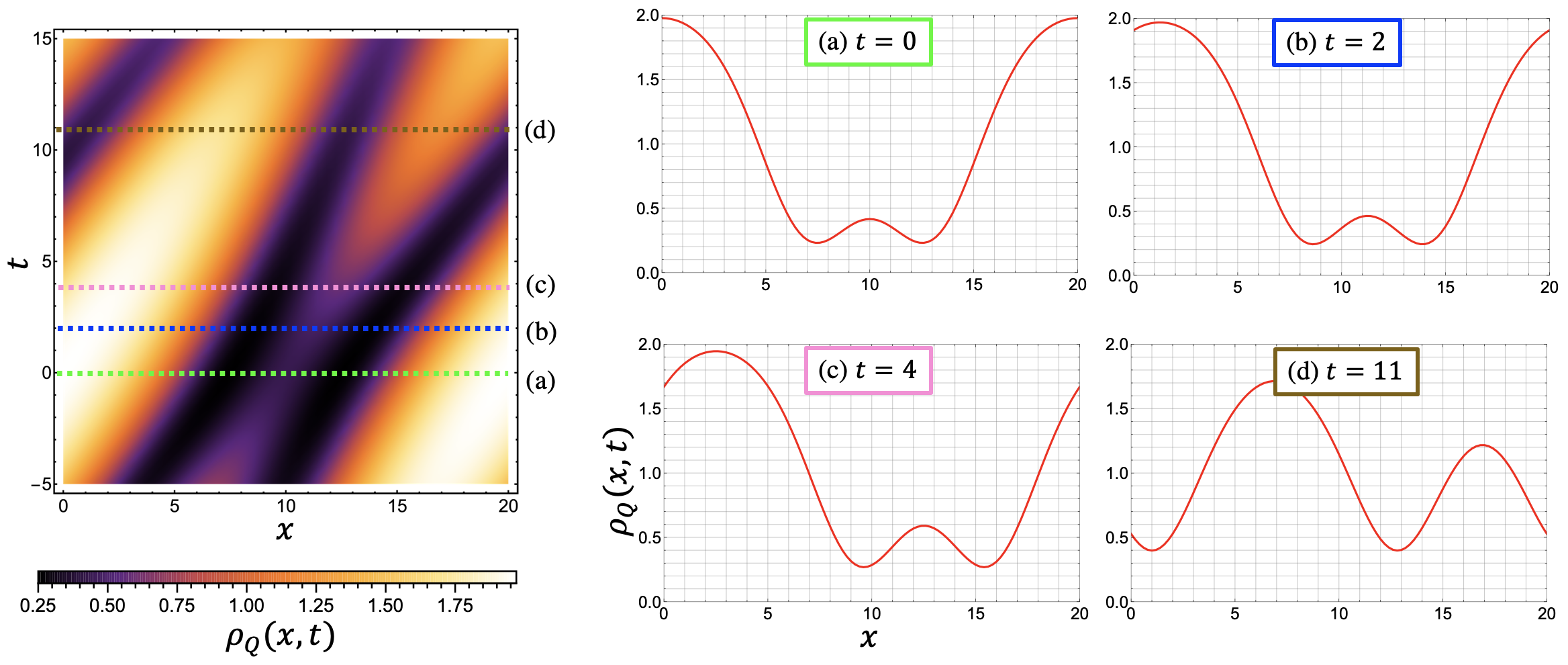}
    \caption{ (Color online) \Add{2D density plot of} \Erase{Time evolution} local density \Erase{profile} $\rho_Q(x,t)$ 
    \Add{in the space-time diagram} (left panel) and  snapshots 
    \Add{of the density profile of $\rho_Q(x,t)$} at $t=0,\ 2,\ 4$ and 11 (right panels),  for $N=L=20$ \Erase{$X_1=X_2=0$,} and $c=0.05$. \Add{The corresponding quantum state is given by Equation (\ref{eq:XN}) with $X_1=X_2=0$.} \Erase{Panel (a) shows the density profile at $t=0$, which corresponds to the green dotted line in the left panel of the density plot.} 
    Panels (a), (b), (c),  and (d) show the density profile at $t=0,\ 2,\ 4$, and 11, which correspond to the green, blue, pink, and ocher dotted lines in the \Add{2D density} plot \Add{(left panel)}, respectively. The frames of the legends of panels (a), (b), (c), and (d) correspond to the green, blue, pink, and ocher colors, respectively.}
    \label{fig:dynamics_N20_c005_den_sameposition}
\end{figure}

\begin{figure}[ht]
    \centering
    \includegraphics[width=0.83\linewidth]{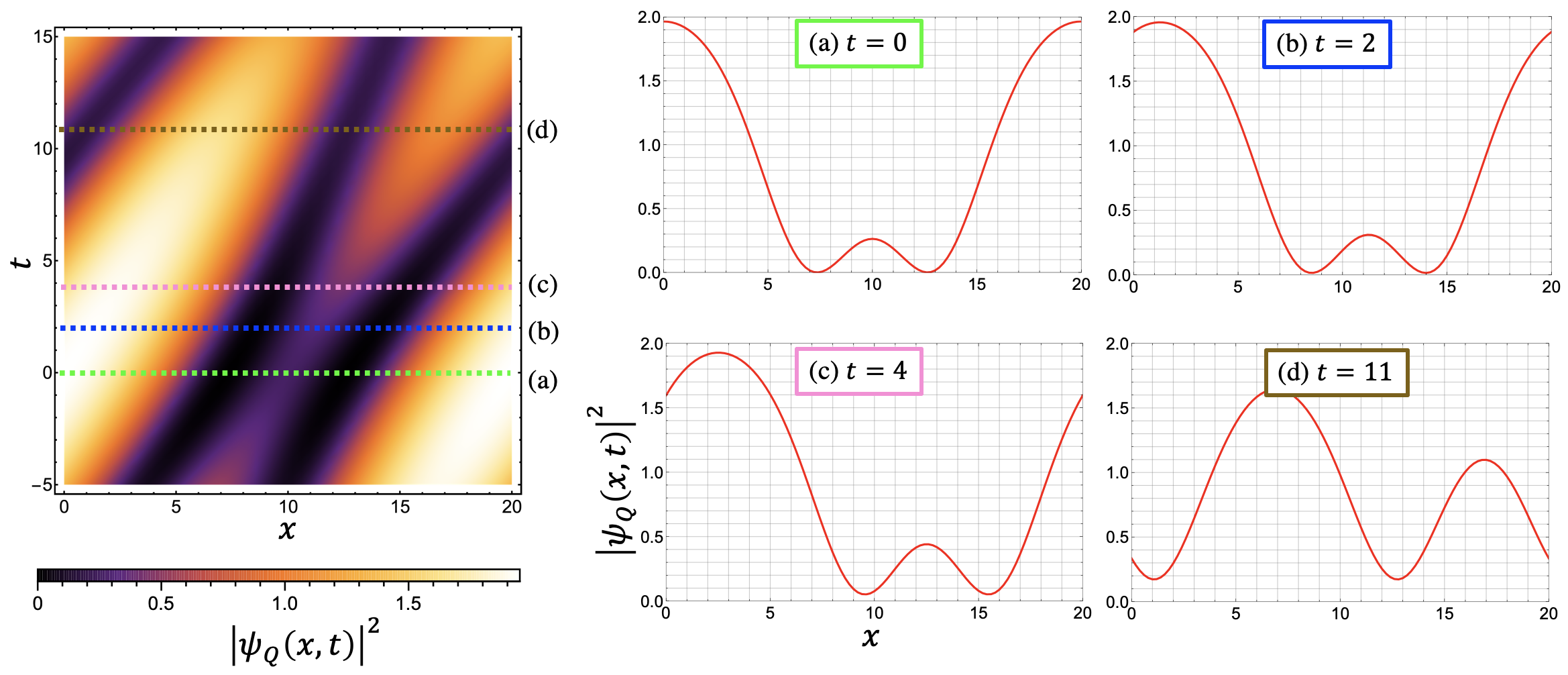}
    \caption{ (Color online) \Add{2D density plot of} \Erase{Time evolution of} the square amplitude \Add{$|\psi_Q(x,t)|^2$ }
    of the matrix element $\psi_Q(x,t)$ \Add{(left panel) and snapshots of the profile of the square amplitude (i.e., $|\psi_Q(x,t)|^2$ versus $x$) at $t =0, 2, 4$ and 11 (right panels)}, for $N=L=20$ \Erase{$X_1= X_2=0$,} and $c=0.05$. \Add{The corresponding quantum states are given by Equation (\ref{eq:XN}) with $X_1=X_2=0$ for $N-1$ and $N$ particles.} Panel (a) shows the density profile at $t=0$, which corresponds to the green dotted line of the \Add{2D} density plot in the left panel. Panels (b), (c), and (d) show the density profile at $t=2, 4$, \Add{and} 11, which correspond to the blue, pink, and ocher dotted lines  of the \Add{2D} density plot in the left panel, respectively. The frames of the legends of panels (a), (b), (c), and (d) correspond to the green, blue, pink, and ocher colors, respectively.}
    \label{fig:dynamics_N20_c005_field_samepos}
\end{figure}

By \Add{placing the positions of the notches for the quantum dark-solitons} \Erase{putting the quantum soliton positions} $X_1$ and $X_2$ at the same \Add{point} \Erase{positions}, the \Erase{resultant} profiles of the density and square amplitude \Add{derived in time evolution} are plotted in \Add{Figures} \ref{fig:dynamics_N20_c005_den_sameposition} and \ref{fig:dynamics_N20_c005_field_samepos}, respectively. 
\Erase{For both cases,} \Add{In both profiles of the density and the square amplitude} \Add{it seems as if} the \Add{two} notches repel each other in \Add{time} evolution.

The quantum \Add{double dark-}soliton \Add{state} with \Erase{two}\Add{overlapping positions of two notches} has different properties 
\Erase{for} \Add{in the profiles of} the density and the square amplitude from \Erase{those of} the quantum \Add{double dark-}soliton state in \Eref{eq:XN} \Erase{where} \Add{with different initial positions of two notches as} $X_1=\frac{L}{4}$ and $X_2=\frac{3L}{4}$. \Erase{For} \Add{In} the density profile of \Fref{fig:dynamics_N20_c005_den_sameposition}
\Erase{case, we observed}
the notches are much deeper than those in \Fref{fig:dynamics_N20_c005_den}\Add{, similarly as the notches in \Fref{fig:dynamics_N20_c005_field_samepos} of the square amplitude profile}. 
\Erase{For} \Add{In the profile of} the square amplitude \Erase{case}, the values at the bottoms of the notches increase in time evolution: The values of the square amplitude at the bottoms of notches are not close to zero \Erase{any longer} at $t=11$ in \Fref{fig:dynamics_N20_c005_field_samepos}.
Thus, for the quantum \Add{double dark-}soliton with two overlapping positions \Add{of notches} the difference between the density profile and the square amplitude \Add{profile} is smaller than in \Fref{fig:dynamics_N20_c005_den} and \Fref{fig:dynamics_N20_c005_field}.

\begin{figure}[ht]
    \centering
    \includegraphics[width=0.6\linewidth]{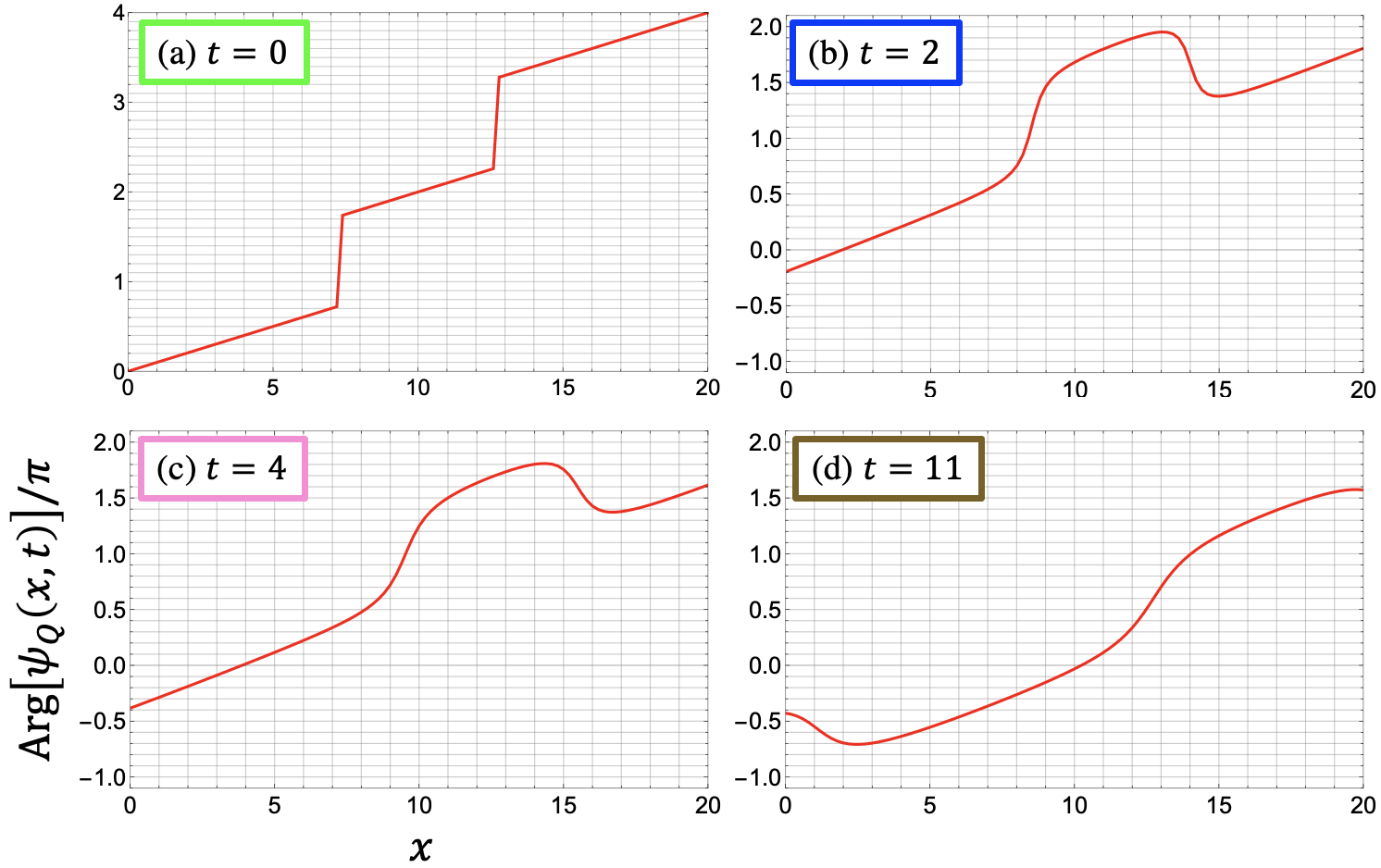}
    \caption{ (Color online) \Erase{The} Snapshots \Add{in} \Erase{of} the time evolution of the phase profile for the matrix element $\psi_Q(x,t)$ at $t=0, 2, 4$ and 11 for $N=L=20$ \Erase{$, X_1= X_2=0$,} and $c=0.05$. \Add{The corresponding quantum states are given by Equation (\ref{eq:XN}) with $X_1=X_2=0$ for $N-1$ and $N$ particles.}}
    \label{fig:N20_c005_ph_sameposition}
\end{figure}

\Add{The snapshots of} the phase profile in time evolution 
are exhibited in \Fref{fig:N20_c005_ph_sameposition} for the quantum 
 \Add{double dark-}soliton state with two overlapping positions \Add{of notches}. \Erase{For} \Add{In} each panel, the phase profile satisfies the boundary condition \Add{specified by} \Erase{with}  a nonzero winding number $J$: $\textrm{Arg}[\psi_Q(x+L,t)]=\textrm{Arg}[\psi_Q(x,t)]+2\pi J$ \Erase{where winding number} ($J\in \mathbb{Z}$). At $t=0$, the winding number \Erase{of the phase} was given by $J=2$, while it suddenly changed to $J=1$ at $t=0.05$. After the change of the winding number, the phase profile became smoother in shape gradually in time evolution. Furthermore, \Add{we observe in} \Fref{fig:N20_c005_ph_sameposition} \Erase{shows} that the whole \Add{phase profile} was shifted toward the negative direction \Add{step-by-step} in time evolution. 
 It is also the case in \Fref{fig:N20_c005_uni_Phase_time}\Add{:} \Erase{representing} 
 \Add{The whole phase profile was shifted in the negative direction for the quantum double dark-soliton state with initial positions of notches placed  \Erase{put in the same distance} at $X_1=\frac{L}{4}$ and $X_2=\frac{3L}{4}$.} 

\Add{The abrupt change of the winding number may occur in time evolution for the phase profile associated with the quantum states, i.e., the phase profile of the matrix element of the field operator between the quantum double dark-soliton states in \Eref{eq:time_matele}.}  
The boundary condition of the phase \Add{is given by } 
the form of \Eref{eq:pbc_winding} \Add{for solutions of the GP equations and 
also for the phase profile associated with the quantum states in \Eref{eq:time_matele}}. 
However, \Erase{the time evolution of} 
the quantum states do not depend on the boundary conditions of classical solutions. 
\Add{It is sufficient if the phase profile associated with the quantum states satisfies one of the boundary conditions of \Eref{eq:pbc_winding} specified by an integer $J$, which we have called the winding number. Thus, the winding number $J$ may change abruptly in time evolution in the phase profile associated with the quantum double dark-soliton states in \Eref{eq:time_matele}.     
}
\Erase{Since we evaluate the time evolution of the quantum dark-soliton state exactly, the abrupt change of the winding number may occur for the quantum dark-soliton state.}


\subsection{Time evolution of quantum double dark-soliton state with the ideal Gaussian weights}\label{subsec:qsol_g}
Let us consider the Gaussian \Add{weighted} superposition of the excited states consisting of two particle-hole excitations which are determined by a pair of holes $\bm{p} = \{p_1, p_2\}$ in the set $\bm{P}$:
\begin{eqnarray}
    |X_{1}, X_{2}, N\ket_G =\frac{1}{\sqrt{\mathcal{N}}} \sum_{\bm{p}\in \bm{P}} G_{P_0\Add{\sigma_0}}(p_1) G_{P'_0\Add{\sigma'_0}}(p_2) e^{i (p_{1} X_{1}+ p_{2} X_{2})}\left|p_{1}, p_{2}, N\right\rangle.
    \label{eq:2quantum_soliton_state_g}
\end{eqnarray}
Here, $\mathcal{N}$ is a normalization factor and 
the set $\bm{P}$ is the same as given in Section \ref{subsec:qdoublesolstate}. The Gaussian function is given by
\begin{eqnarray}
    G_{P\Add{\sigma}}(q)= \exp \left[-\frac{\left(q-P\right)^{2}}{4 \sigma^2}\right] 
    \label{eq:Gaussian_weight} 
\end{eqnarray}
with two Gaussian parameters $(P,\sigma)$ \cite{kaminishi2020construction}. The parameters $P$ and $\sigma$ are determined by the target soliton depth $d$ and the density $n=N/L$:
\begin{eqnarray}
    P(d) &= 2 n \left\{ \frac{\pi}{2}-\left[ \sqrt{\frac{d}{n}\left(1-\frac{d}{n}\right)} +\arcsin{\left(\sqrt{\frac{d}{n}}\right)} \right] \right\} \, ,  \label{eq:g_param_p}
    \\
    \sigma^2 \Add{(d)}& =\frac{4}{3} n \sqrt{n c}\left( 1-\frac{d}{n}\right) ^\frac{3}{2} \label{eq:g_param_sigsq}.
\end{eqnarray}
\Add{Here we have defined the soliton depth $d$ by the smallest value in the density profile of a single dark-soliton. It is different from the ``depth of a notch'' defined in Section 3.1.1.} The target soliton depth $d$ is expressed with the dark soliton solution to the GP equation moving with \Erase{speed}
\Add{velocity} $v$ in the thermodynamic limit $\phi_P^\infty(x)$ \cite{kaminishi2020construction}:
\begin{eqnarray}
\sqrt{d} =|\phi_P^\infty(x=0)|=\sqrt{n}\left|\frac{v}{v_{c , \infty}} \right|.
\label{eq: target_depth}
\end{eqnarray}
Here $|\phi_P^\infty(x=0)|$ denotes the square root 
of the local density at the origin, \Erase{i.e., at} \Add{which is} the position of \Erase{a} \Add{the} notch in the thermodynamic limit, and $v_{c,\infty}$ is called the critical velocity \Add{of the infinite system. When the system size $L$ is finite, the largest velocity of the elliptic dark-soliton solution of the GP equation 
is denoted by the critical velocity $v_{c}$\cite{kinjo2022quantum}. 
It approaches the critical value $v_{c,\infty}$ in the limit of 
sending the system size $L$ to $\infty$. }

\Erase{The density profile and the square amplitude profile of the matrix element} 
\Add{The exact profiles in time evolution are numerically derived for the local density $\rho_Q(x,t)$ and the square amplitude of the matrix element $\psi_Q(x,t)$ of the field operator} 
\Erase{evolve in time} 
by \Add{calculating} \Erase{applying} the time-dependent \Add{matrix elements of the} field operator 
\Add{between the Gaussian weighted quantum states of  
 Equation (\ref{eq:2quantum_soliton_state_g})}, 
 similarly as we have \Erase{done} \Add{demonstrated in \Eref{eq:time_den} and \Eref{eq:time_matele}} of Section \ref{subsec:timeevolve_qsol_wEqualweight} \Add{for the quantum double dark-soliton state constructed with equal weight}. 
\Add{For the Gaussian weighted quantum double dark-soliton state,} by assigning \Add{ a pair of proper values of the} target soliton depth \Add{$d$} to the two notches \Add{of a given superposition of quantum states of Equation (\ref{eq:2quantum_soliton_state_g})}, we \Add{can construct} \Erase{have constructed} 
a quantum double dark-soliton state such that its density profile has two distinct notches with different depths. 

\begin{figure}[ht]
    \centering
    \includegraphics[width=0.83\linewidth]{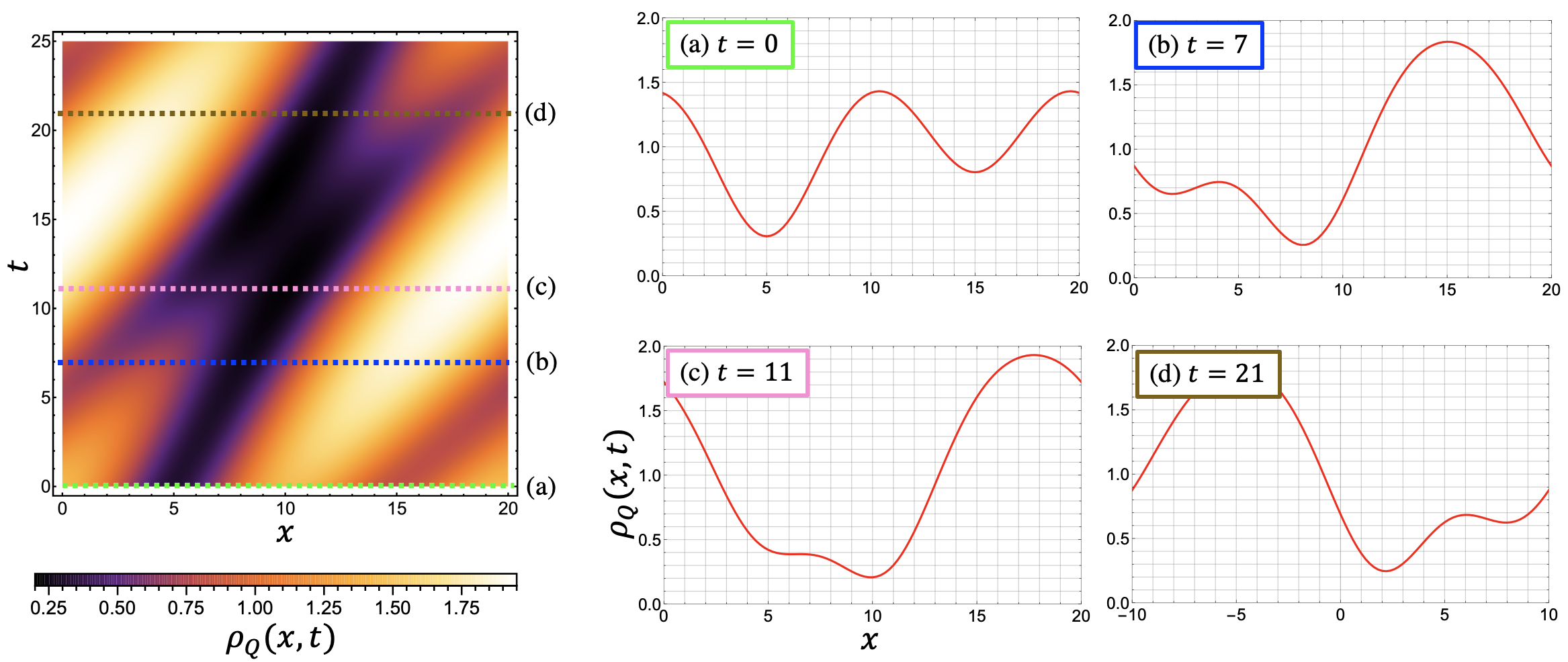}
    \caption{ (Color online) \Erase{Time evolution of} \Add{2D density plot of local density $\rho_Q(x,t)$ in the space-time diagram (left panel) and the snapshots of the density profile for the quantum double dark-soliton state constructed} with the Gaussian weights for $N=L=20$ \Erase{, $X_1=\frac{L}{4}, X_2=\frac{3L}{4}$,} and $c=0.05$.
     \Add{The corresponding quantum state is given by Equation (\ref{eq:2quantum_soliton_state_g}) with $X_1=L/4$ and $X_2=3L/4$.}
    Panels (a), (b), (c), and (d) show the density profile at $t=0, 7,11$, and 21, which correspond to the green, blue, pink, and ocher dotted lines in the left panel of the \Add{2D density} \Erase{density} plot, respectively. The frames of the legends of panels (a), (b), (c), and (d) correspond to the green, blue, pink, and ocher colors, respectively.}
    \label{fig:N20_c005_den_G}
\end{figure}

\begin{figure}[ht]
    \centering
    \includegraphics[width=0.83\linewidth]{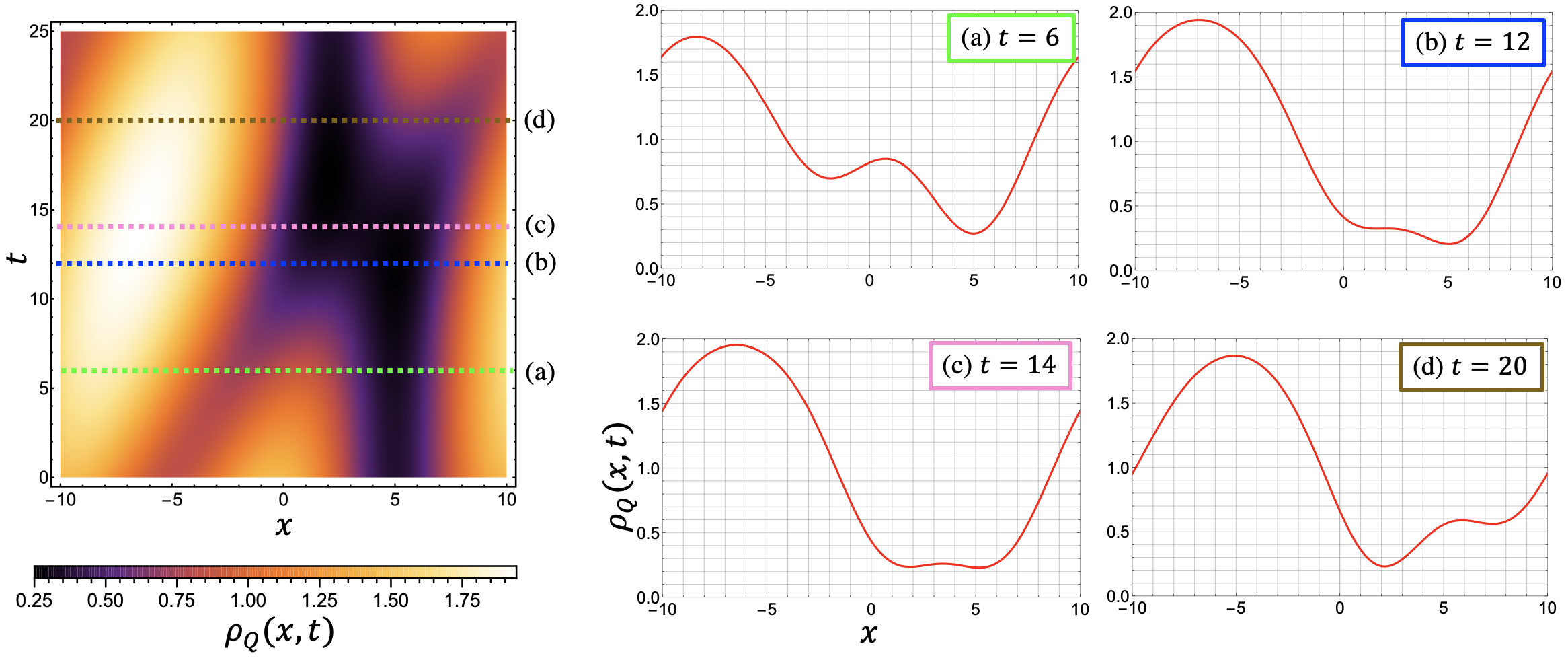}
    \caption{ (Color online) \Add{2D density plot} \Erase{Time evolution} of 
    \Add{local density $\rho_Q(x,t)$ in the space-time diagram derived 
    by the Galilean transformation} 
    \Erase{the density profile}  (left panel) and \Erase{its} 
    \Add{corresponding} snapshots \Add{of the density profile} at $t=6,12,14$, and 20  
    \Add{in the quantum double dark-soliton state constructed with the Gaussian weights}
    for $N=L=20$ \Erase{$, X_1=-L/4,\ X_2=L/4$,} and $c=0.05$.  \Add{The corresponding quantum state is given by Equation (\ref{eq:2quantum_soliton_state_g}) with $X_1=-L/4$ and $X_2=L/4$.} Panels (a), (b), (c), and (d) show the density profile at $t=6,12,14$, 
    \Add{and} 22, which correspond to the green, blue, pink, and ocher dotted lines in the left panel of the \Add{2D} density plot, respectively. The frames of the legends of panels (a), (b), (c), and (d) correspond to the green, blue, pink, and ocher colors, respectively.}
    \label{fig:N20_c005_Gtrans_den_G}
\end{figure}

\begin{figure}[h]
    \centering
    \includegraphics[width=0.83\linewidth]{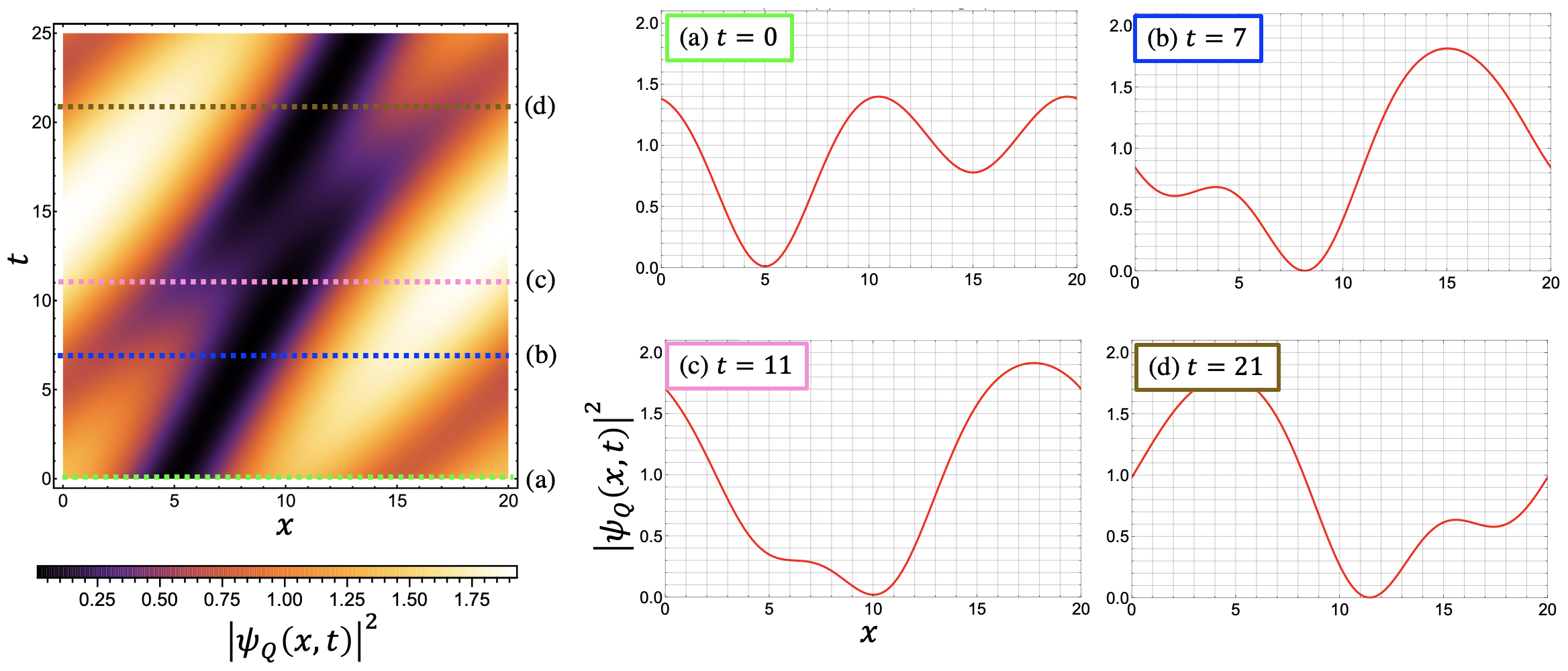}
    \caption{ (Color online) \Add{2D density plot} \Erase{Time evolution} of the square amplitude \Add{of the matrix element $\psi_Q(x,t)$} 
    \Erase{with Gaussian weights and its} \Add{and the }snapshots 
    \Add{of the square amplitude profile} at $t=0, 7, 11$, \Add{and} 21\Add{, in the Gaussian weighted quantum double dark-soliton state} 
    for $N=L=20$ \Erase{, $X_1=\frac{L}{4}, X_2=\frac{3L}{4}$,} and $c=0.05$. \Add{The corresponding quantum states are given by Equation (\ref{eq:2quantum_soliton_state_g}) with $X_1=\frac{L}{4}$ and $X_2=\frac{3L}{4}$ for $N-1$ and $N$ particles.} Panels (a), (b), (c), and (d) show the density profile at $t=0, 7, 11$, \Add{and} 21, which correspond to the green, blue, pink, and ocher dotted lines in the left panel of the \Add{2D} density plot, respectively. The frames of the legends of panels (a), (b), (c), and (d) correspond to the green, blue, pink, and ocher colors, respectively.}
    \label{fig:N20_c005_amp_G}
\end{figure}

\begin{figure}[h]
    \centering
    \includegraphics[width=0.6\linewidth]{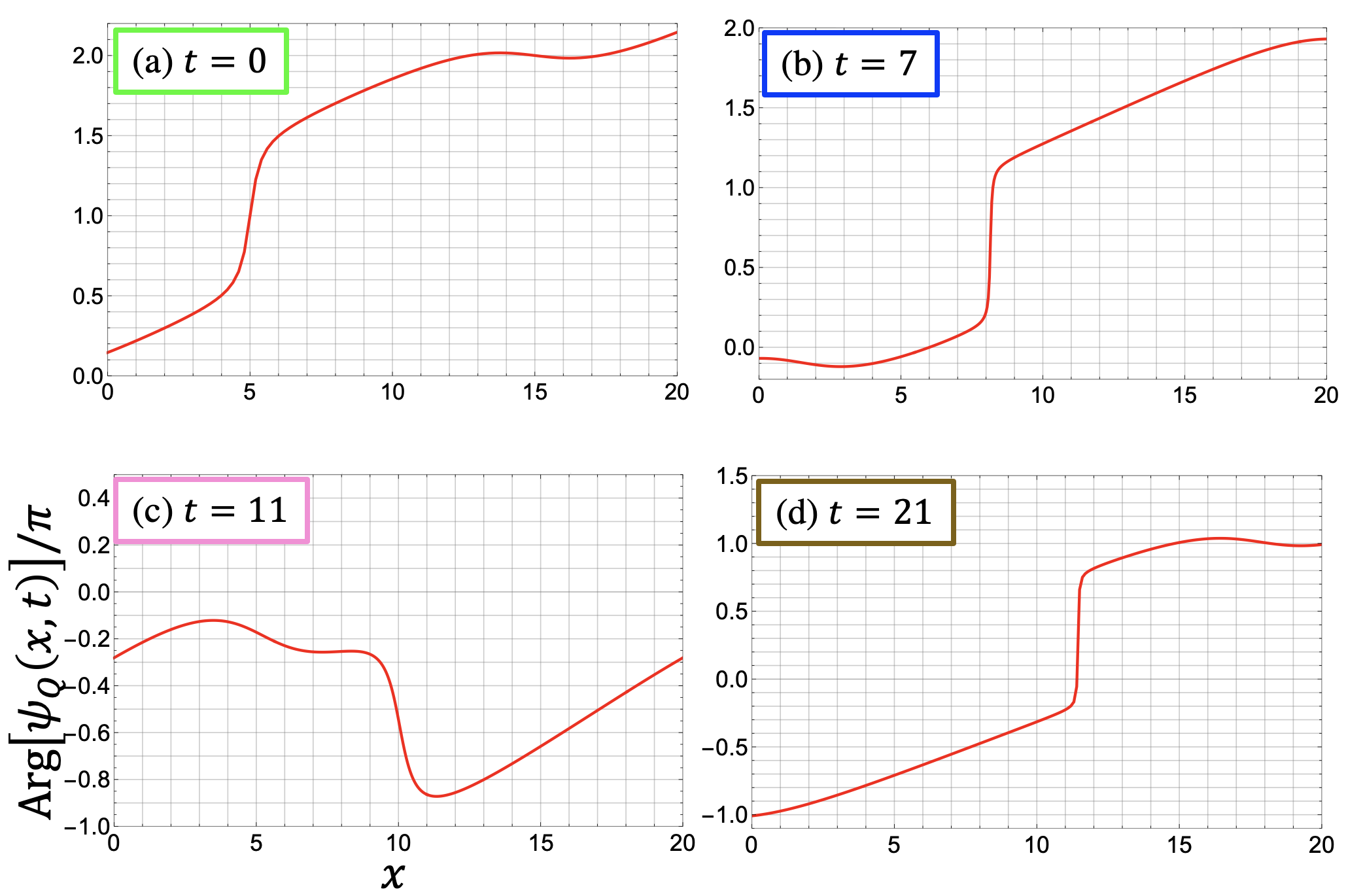}
    \caption{ (Color online) \Erase{The} Snapshots of \Erase{the time evolution of} the phase profile \Add{of the matrix element $\psi_Q(x,t)$} \Erase{with Gaussian weights} at $t=0, 7, 11$, \Add{and} 21 for the Gaussian weighted quantum double dark-soliton states 
    with $N=L=20$ \Erase{,$X_1=\frac{L}{4}, X_2=\frac{3L}{4}$,} and $c=0.05$. 
    \Add{The corresponding quantum states are given by Equation (\ref{eq:2quantum_soliton_state_g}) with $X_1=\frac{L}{4}$ and $X_2=\frac{3L}{4}$ for $N-1$ and $N$ particles.}}
    \label{fig:N20_c005_ph_G}
\end{figure}

\begin{figure}[h]
    \centering
    \includegraphics[width=0.83\linewidth]{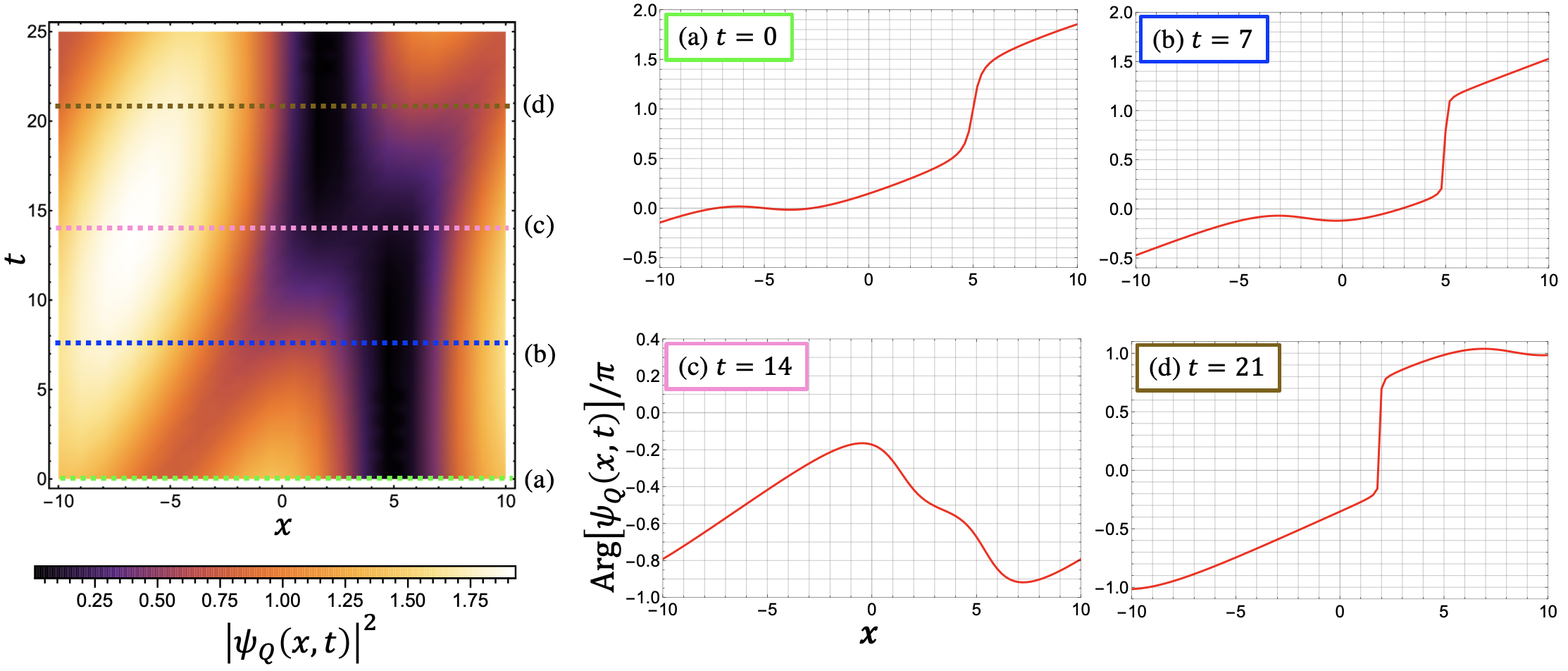}
    \caption{ \Add{(Color online) 2D density plot of the square amplitude profile of the matrix element $\psi_Q(x,t)$ (left panel), and the snapshots of the phase profile for the matrix element $\psi_Q(x,t)$  at $t=0,7,14$, and 21 (right panels), 
    in the Gaussian weighted quantum double dark-soliton state for $N=L=20$ and $c=0.05$, observed in the inertial frame of reference 
    moving with the deeper notch, derived 
    by the Galilean transformation. The corresponding quantum state is given by Equation (\ref{eq:2quantum_soliton_state_g}) with $X_1=-L/4$ and $X_2=L/4$. Panels (a), (b), (c), and (d) show the phase profile at $t=0,7,14,21$, which correspond to the green, blue, pink, and ocher dotted lines in the left panel of the 2D density plot, respectively. The frames of the legends of panels (a), (b), (c), and (d) correspond to the green, blue, pink, and ocher colors, respectively.}}
    \label{fig:N20_c005_amph_Gcoord}
\end{figure}

\begin{figure}[h]
    \centering
    \includegraphics[width=0.6\linewidth]{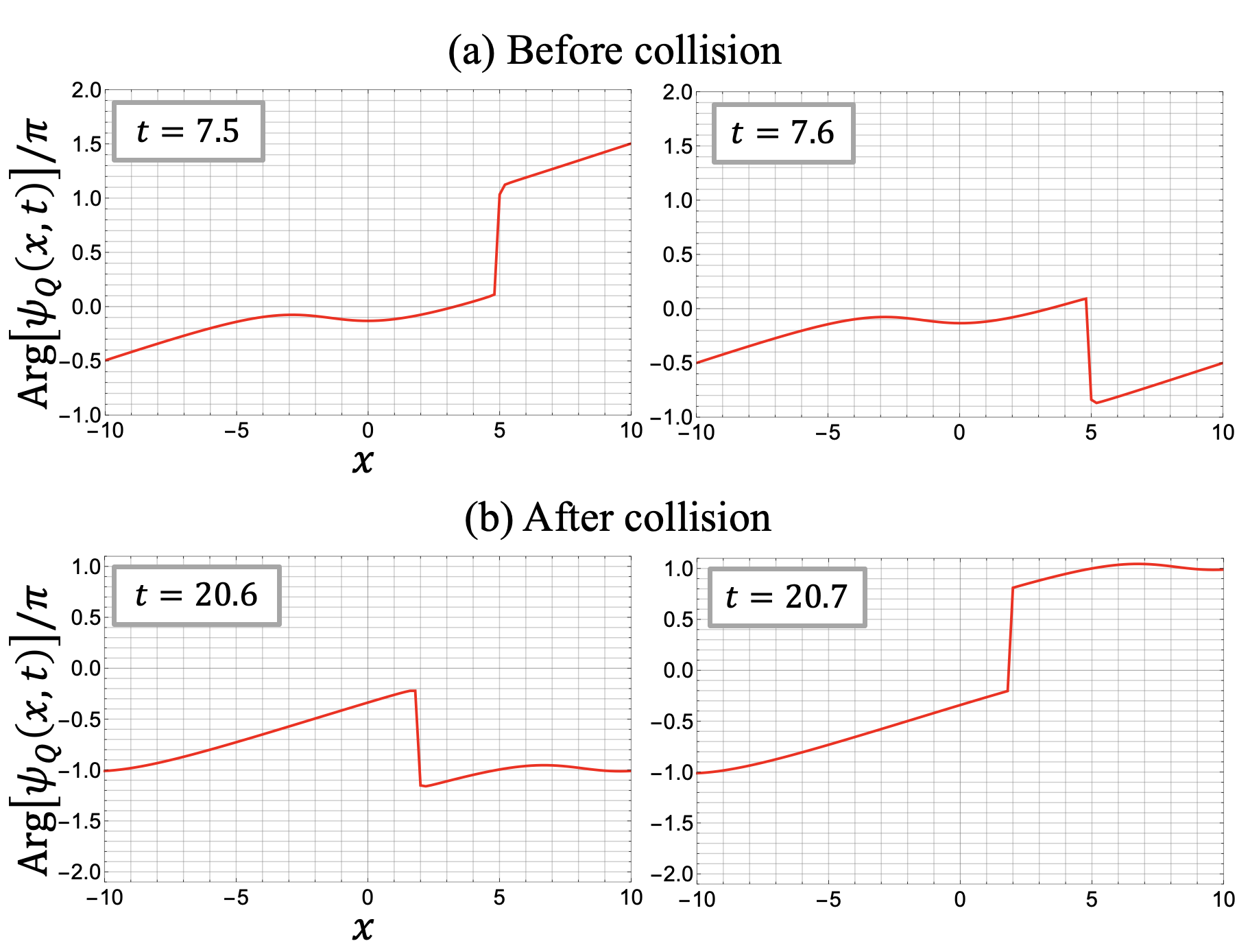}
    \caption{\Add{ (Color online) Snapshots in the time evolution of the phase profile associated with the matrix element $\psi_Q(x,t)$ for 
    the Gaussian weighted quantum double dark-soliton state, observed in the inertial frame of reference derived by the Galilean transformation: (a) before collision and (b) after collision at $t=7.5$ and 7.6,  and $t=20.6$ and 20.7, respectively. We set $N=L=20$ and $c=0.05$. The corresponding quantum state is given by Equation (\ref{eq:2quantum_soliton_state_g}) with $X_1=-L/4$ and $X_2=L/4$.}}
    \label{fig:N20_c005_ph_G_changingW}
\end{figure}

\Add{We have constructed several quantum double dark-soliton states in which the density profile has two distinct notches with different depths.} 
In Figures \ref{fig:N20_c005_den_G}, \ref{fig:N20_c005_Gtrans_den_G}, \ref{fig:N20_c005_amp_G}, \ref{fig:N20_c005_ph_G}, \ref{fig:N20_c005_amph_Gcoord} \Add{and} \ref{fig:N20_c005_ph_G_changingW}
we  set the target soliton depths as $d=0.6$ and $d=0.0$ \Add{to the two notches, respectively}, \Add{and we generated the Gaussian weights by making use of \Eref{eq:Gaussian_weight}.} Here, the \Add{corresponding} Gaussian parameters are given by $(P_0, \sigma) =  (0.124027\pi ,0.106667)$ and $(P'_0, \sigma') = (\pi ,0.421637)$, 
\Add{respectively, } which are\Erase{calculated by} 
\Add{derived by making use of}
 Equations (\ref{eq:g_param_p}) and (\ref{eq:g_param_sigsq}).
\Add{We have thus obtained the quantum double dark-soliton \Add{state} of distinct narrow notches with different depths. 
\Add{Here we recall} that single dark-solitons with different depths
have different speeds in the same direction for the GP equation.} 

We \Erase{thus} observe\Erase{d} the scattering of two notches \Add{in the density and phase profiles of} the quantum double dark-soliton state. 
\Erase{, which} \Add{It exhibits the phase shift which is a characteristic property in soliton-soliton collisions \cite{kaminishi_JPSmtg, PhysRevResearch.4.L032047}, as shown in the density profile. 
\Add{We remark that} the 2D density plot of the local density in the space-time diagram and the snapshots of the density profile at different times are presented in Figure \ref{fig:N20_c005_den_G} for} the quantum double dark-soliton state constructed with the Gaussian weights for $c=0.05$.
As the two \Add{notches of the double} dark-soliton approached \Add{each other}, they moved along approximately straight and linear  trajectories with different constant velocities. 
The collision occurred around at \Add{a time interval including} $t=11$ (see panel (c), which corresponds to the pink dotted line in the left panel of \Fref{fig:N20_c005_den_G}). After the collision, each of the dark solitons travelled \Erase{with} \Add{at} the same \Erase{speed} \Add{velocity} before the collision. \Add{Furthermore,} we confirm \Add{that} the phase shift 
\Add{occurred} after the collision \Add{in the left panel of 
Figure \ref{fig:N20_c005_den_G}}.

\Erase{Figure \ref{fig:N20_c005_den_G} shows the time evolution of the density profile of the quantum state constructed with the Gaussian weights for $c=0.05$. 
the quantum state constructed with the Gaussian weights for $c=0.05$.
We have constructed the double dark-solitons of distinct narrow notches with different depths. Note that different depths of
single dark-solitons give different speeds in the same direction. 
We thus observed the scattering of two notches in the quantum double dark-soliton state, which has a characteristic property in soliton-soliton collisions \cite{kaminishi_JPSmtg}. As the two dark-solitons approached, they moved along approximately straight line trajectories with different constant velocities. The collision occurred around at $t=11$ (see panel (c), which corresponds to the pink dotted line in the left panel of \Fref{fig:N20_c005_den_G}). After the collision, each of the dark solitons travelled with the same \Erase{speed} \Add{velocity} before the collision. We confirmed the phase shift after the collision.}


Let us investigate the phase shift \Erase{more} explicitly.
By applying the Galilean transformation, that is, \Add{in 
\Fref{fig:N20_c005_Gtrans_den_G} we observe the scattering process in the inertial frame of reference} moving with the \Add{left-hand-side notch of the } \Erase{coordinate of the left-side} quantum \Add{
double} dark-soliton in Figure \ref{fig:N20_c005_den_G}. We clearly confirm the phase shift after the collision as shown in \Fref{fig:N20_c005_Gtrans_den_G}. The position of the deeper \Add{notch} \Erase{soliton} is shifted from $x=5$ to $x=2$ as shown in panel (d) of \Fref{fig:N20_c005_Gtrans_den_G}. \Add{The position shift of the deeper notch} corresponds to the phase shift due to the collision of the two dark-solitons.

\Fref{fig:N20_c005_amp_G} shows the time evolution of the square amplitude \Add{profile} of the matrix element of the field operator \Add{for the quantum double dark-soliton states constructed} with the Gaussian weights for $c=0.05$. The quantum state is the same as that of \Fref{fig:N20_c005_den_G}. We have constructed the double dark-solitons of distinct narrow notches with different depths \Add{not only in the density profile but also in the square amplitude profile, i.e., the graph of $|\psi_Q(x,t)|^2$ versus $x$.}
\Erase{as well as in the density profile.} 
We observe the scattering of two notches in the quantum double dark-soliton states. As the two \Add{notches of the double dark-soliton states} approached \Add{each other}, they moved along approximately straight and linear trajectories with different constant velocities, as shown in \Fref{fig:N20_c005_amp_G}. The collision occurred 
\Erase{around at} \Add{in the time interval including} $t=11$ (see panel (c), which corresponds to the pink dotted line in the left panel of \Fref{fig:N20_c005_amp_G}). After the collision, each of the dark solitons travelled with the same \Erase{speed} \Add{velocity} before the collision. We observe\Erase{d} at least approximately the same phase shift as shown in \Fref{fig:N20_c005_den_G}.
\Add{We remark that panel (a) of \Fref{fig:N20_c005_amp_G}, the square amplitude \Add{profile} of the matrix element 
for the quantum states constructed with the Gaussian weights, corresponds to the panel of $c=0.05$ in Figure 13 of Ref. \cite{kinjo2022quantum}.}

We now \Erase{show} \Add{demonstrate}
that the winding number \Add{changed} during the scattering process \Add{in the time evolution of the Gaussian weighted quantum double dark-soliton states}. \Fref{fig:N20_c005_ph_G} shows the time evolution of the phase profile. In each panel, the phase profile satisfies the boundary condition: $\textrm{Arg}[\phi_Q(x+L,t)]=\textrm{Arg}[\phi_Q(x,t)]+2\pi J$ with a winding number $J$.
At the \Erase{beginning} \Add{initial time} $t=0$, \Erase{when} the \Add{two notches of the} quantum dark-soliton were \Erase{put in the same distance} \Add{located at the most distant points from each other such as $X_1=L/4$ and $X_2=3L/4$, and} 
the winding number is \Add{given by} $J=1$. When the two \Add{notches of the} quantum dark-soliton states \Erase{were} \Add{became} very close \Erase{enough} \Add{in space}, the winding number was suddenly changed to $J=0$, \Erase{approximately at} \Add{in the time interval including} $t=11$, as shown in panel (c) of \Fref{fig:N20_c005_ph_G}. After the collision, the winding number was recovered: The winding number at $t=21$ \Erase{became} 
\Add{was given by} $J=1$, as shown in panel (d) of \Fref{fig:N20_c005_ph_G}. %
\Add{We remark that panel (a) of \Fref{fig:N20_c005_ph_G}, the phase profile of the matrix element between the Gaussian weighted quantum double dark-soliton states, corresponds to the panel of $c=0.05$ in Figure 14 of Ref. \cite{kinjo2022quantum}.}

\Add{We explicitly evaluate the phase shift due to the scattering of two notches in the quantum double dark-soliton state. The left panel of Figure \ref{fig:N20_c005_amph_Gcoord} shows the square amplitude profile of the matrix element $\psi_Q(x, t)$ in time evolution observed in the inertial frame of reference moving together with the deeper notch of the quantum double dark-soliton. The abrupt increase (or decrease) in the phase profile, which we call a phase jump,  was located at the position of the deeper notch of the double dark-soliton, as shown in panels (a), (b), and (d) of Figure \ref{fig:N20_c005_amph_Gcoord}: It was located at $x=5$ in panels (a) and (b), and at $x=2$ in panel (d). Thus, the position of the deeper notch in the double dark-soliton was shifted after the collision in the inertial frame of reference. It corresponds to the phase shift due to the scattering of the two notches. 
}

\Add{Let us investigate the changes of the winding number in time evolution in detail. The winding number $J$ was equal to zero when the two notches of the quantum double dark-soliton were close to each other in space, as shown in panel (c) of Figure  \ref{fig:N20_c005_amph_Gcoord}. Figure \ref{fig:N20_c005_ph_G_changingW} exhibits that the abrupt changes of the winding number $J$ from 1 to 0 and from 0 to 1 occurred before and after the process of the soliton collision.  
At time $t=7.5$ the value of the phase abruptly increases in space 
with respect to the $x$ coordinate at the position $x=5$ of the phase jump, while at time $t=7.6$ it suddenly decreases in space at the position $x=5$ of the phase jump. Similarly, the corresponding change of the winding number occurred during the time interval between $t=20.6$ and $20.7$ after the collision. 
}

\Add{We recall that it is not necessary for the winding number in the phase profile of a quantum state to be conserved during the time evolution of the quantum system.  The winding number is defined for the corresponding classical system, i.e., the GP equation, or for the phase profile of the quantum system.  
The dynamics of the quantum system can be much more complex than the 
solutions of the GP equation. When the two notches are far from each other in space, the phase profile of the quantum system is similar to that of the classical solution, while it is not the case when they collide with each other since they are very close in space.  
} 

\Erase{The} \Add{In summary, the Gaussian weighted superposition of the two-hole excited states} has lead to the quantum double dark-soliton states in which  
\Add{two notches have} different depths \cite{kinjo2022quantum}. 
\Erase{This gives} \Add{It follows that} the notches of the quantum double dark-soliton state have different velocities, and hence we have observed the scattering of two notches in the quantum double dark-soliton state exactly. We have also shown that the winding number of a quantum double dark-soliton state changed when the two notches approach each other\Add{, explicitly for the Gaussian weighted quantum double dark-soliton states}. 

\Add{
We remark that one can make the quantum single dark-soliton black by making use of the Gaussian weights, as shown in Ref. \cite{kaminishi2020construction}. However, for the quantum double dark-soliton, it seems that it is difficult to construct the double black-soliton only by applying the Gaussian weights to the superposition of a set of two-hole excitations. }

\newpage

%
\section{Finite-Size Scaling of BEC}

\subsection{Motivation to study the quasi-BEC in 1D \Add{for the ground state}}
In 1D systems quantum fluctuations 
play a key role and often give subtle and nontrivial effects. 
It is known that BEC occurs even 
for bosons with repulsive interactions due to the quantum statistical effect among identical particles \cite{leggett2006quantum}.
In fact, the existence of BEC has been proven rigorously 
for interacting bosons confined in dimensions greater than one 
\cite{PhysRevLett.88.170409}. In 1D case there is no BEC for  
bosons with repulsive interactions due to strong quantum fluctuations 
if we \Add{assume} the standard thermodynamic limit with fixed coupling constant \cite{pitaevskii1991uncertainty}.  
On the other hand, if the coupling constant is very weak, 
we may expect that even the 1D bosons with a  
 large but finite number of particles 
undergo a quasi-condensation in which 
 ``a macroscopic number of particles occupy 
a single one-particle state'' \cite{leggett2006quantum}. 
\Add{We call it a quasi-BEC by following the Penrose and Onsager criterion.}

However, it has not been shown explicitly how such a quasi-condensation  
 occurs in interacting bosons in one dimension. Furthermore, it is nontrivial to expect it  for the 1D Bose gas that is solvable by the Bethe ansatz. 
No pair of particles can have the same quasi-momentum in common for a Bethe-ansatz solution. Here we recall that we call the 1D system of bosons interacting with repulsive delta-function potentials the 1D Bose gas. For the impenetrable 1D Bose gas where  
the coupling constant is taken to infinity, 
condensate fractions are analytically and 
numerically studied \cite{PhysRevA.67.043607}, while in the weak coupling case 
it is nontrivial to evaluate the fractions in the 1D Bose gas. 

\Add{We thus study in section 4 how the condensation fraction $n_0$,  i.e., the degree of the quasi-BEC, explicitly depends on the system size $L$, the number of particles $N$ and the coupling constant $c$ in the ground state of the LL model and particularly in the weak coupling case. It will be an illustrative example. 
}

\subsection{Onsager-Penrose criterion of BEC}\label{subsec:criterion_BEC}
Let us review the definition of BEC 
through the one-particle reduced density matrix 
for a quantum system \cite{leggett2006quantum, PhysRev.104.576}. 
 We assume that the number of particles $N$ is very large but finite.  
At zero temperature, the density matrix is given 
by $\hat{\rho}=|\lam\ket\bra \lam |$, 
where $|\lam\ket$ denotes the ground state of the quantum system. 
We define the one-particle reduced density matrix by 
the partial trace of the density matrix with respect to other 
degrees of freedom: $\hat{\rho}_1 = N\rm{tr}_{23\cd N}\hat{\rho}$. 
This matrix is positive definite and hence it is diagonalized as 
\begin{eqnarray}
\hat{\rho}_1=N_0|\Psi_0\ket\bra\Psi_0|+N_1|\Psi_1\ket\bra\Psi_1|
+\cd. 
\end{eqnarray}
Here we put eigenvalues $N_j$ in descending order:   
$N_0 \geq N_1 \geq N_2 \geq \cd>0$.  The sum of all the eigenvalues is given by the number of particles: $\sum_j N_j=N$. 
Here we recall ${\rm{tr}}_1\hat{\rho}_1=N$ due to the normalization: 
$\rm{tr}_{123\cd N} \hat{\rho} =1$.  
Let us denote by $n_0$ the ratio of 
the largest eigenvalue $N_0$ to  particle number $N$: 
\begin{eqnarray} 
n_0:= N_0/N .  
\end{eqnarray}
The criterion of BEC due to Penrose and Onsager \cite{PhysRev.104.576} is given 
as follows: If the largest eigenvalue $N_0$ is of order $N$, i.e., 
 the ratio $n_0$ is nonzero and finite for large $N$,  
then we say that the system exhibits BEC, and  
we call $n_0$ the condensate fraction.  Here  
we also define fractions $n_j$ by $n_j=N_j/N$ for $j=1, 2, \ldots$.

\subsection{Saturation rate \Add{in the form factor expansion at the ground state}} \label{subsec:saturation_rate}

\begin{table}[ht]
\begin{center} \begin{tabular}{cccc}
$c$  &    0.01   &  1 &  100 \\
\hline 
1p1h   & 0.999984   & 0.971538 & 0.693620  \\
2p2h     & $1.59454 \times 10^{-5}$   & 0.0280102 & 0.289056   \\
$n_{\rm sat}$     & 1.00000    & 0.999548 & 0.982676 \\
\hline 
\end{tabular}
\caption{ Fraction $n_{\rm sat}$ of the reduced density operator 
at the origin, $\rho_1(0, 0)$, to the density $n$, 
evaluated by taking the sum over a large number of eigenstates 
$|\mu \rangle$ with 
one particle and one hole (1p1h) or with two particles 
and two holes (2p2h) for $N=L=50$ ($n=1$):   
$n_{\rm sat} = 
\left(  \sum_{\mu}^{1p1h} + \sum_{\mu}^{2p2h}  \right) 
| \bra \mu |\hat{\psi}(0)| \lam \ket |^2 /n . $}
\label{tab:saturation_rate}
\end{center} 
\end{table}

Numerically we  calculate correlation function \Add{in Equation} (\ref{eq:sum-ff}) 
by taking the sum over a large number of eigenstates 
with one particle and one hole (1p1h) and 
those with two particles and two holes (2p2h). 
In order to confirm the validity of the restricted sum, 
we have estimated the ratio 
of the one-particle reduced density operator at the origin 
to density $n$,  $\rho_1(0, 0)/n$,  
through the form factor expansion \Add{in Equation} (\ref{eq:sum-ff})  
for the excitations with 1p1h or 2p2h.  
We express it by $n_{\rm sat}$. 
The estimates of $n_{\rm sat}$ are listed in Table \ref{tab:saturation_rate}.  
The graph of $n_{\rm sat}$ approaches 1 for small coupling constant $c$, 
while it is larger than 0.98 for any value of $c$ in the case of $N=50$.

\subsection{Evaluation of the one-particle reduced density matrix \Add{of the ground state}}

For the LL model, the eigenfunctions of the one-particle reduced density matrix are given by plane waves for any nonzero and finite value of $c$. It is a consequence of the translational invariance of the Hamiltonian of the LL model. We thus have  
\begin{eqnarray}
\rho_1(x,y)
&=\frac{N_0}L+\sum_{j=1}^{\infty} \frac{2N_j}L \cos\left[2\pi j (x-y)/L\right] \, . 
\end{eqnarray} 
\par \noindent 
The eigenvalues of the one-particle reduced density matrix, 
$N_j$, are expressed in terms of the form factor expansion.  
We consider the sum over all the form factors between the ground state, 
$| \lambda \rangle$, and such eigenstates, $| \mu \rangle$,  
that have a given momentum $P_j$ as 
\begin{eqnarray}
N_j
=L\sum_{\mu: P_\mu=P_j}
| \bra \mu |\hat{\psi}(0)| \lam \ket |^2 \, .  \label{eq:Nj}
\end{eqnarray}
In the LL model we have $P_j:=(2\pi/L)j$.

Solving the Bethe ansatz equations for a large number of eigenstates 
we observe numerically that eigenvalues $N_j$ are given in decreasing order with respect to integer $j$: $N_0>N_1>N_2>\cd$. 
It thus follows that condensate fraction 
which corresponds to the largest eigenvalue of 
the one-particle reduced density matrix ${\hat \rho}_1$ 
is indeed given by $n_0=N_0/N$, 
where $N_0$ has been defined by \Add{the} sum \Add{of} \Add{Equation} (\ref{eq:Nj}) 
over all eigenstates with zero momentum.

\subsection{Condensate fraction in the weak coupling regime}

\begin{figure}[ht]
\begin{center}
    \includegraphics[width=0.6\textwidth]{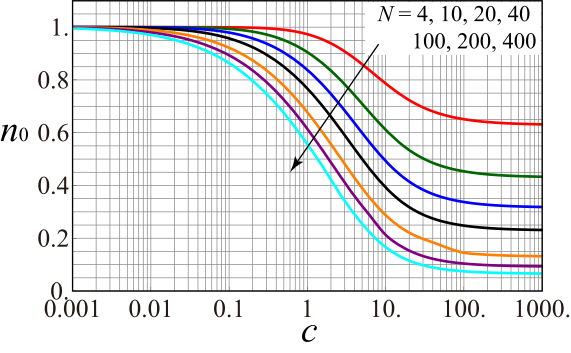}
    \includegraphics[width=0.6\textwidth]{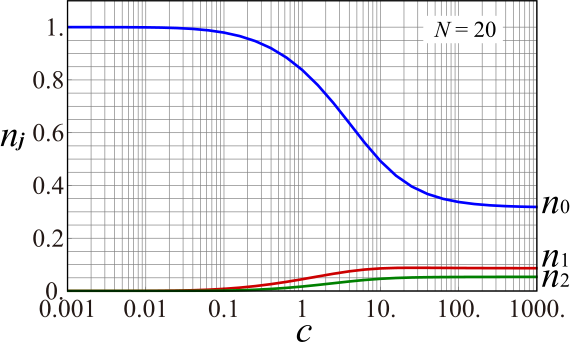}
\caption{(Color online) 
Dependence of fractions $n_j$ on coupling constant $c$. 
In the upper panel: condensate fraction $n_0$ is plotted against  
coupling constant $c$ 
for $N=4$, 10, 20, 40, 100, 200 and 400, 
from the top to the bottom, in red, green, blue, black, orange, purple 
and cyan lines, respectively. 
In the lower panel:  condensate fraction 
$n_0$,  fractions $n_1$ and $n_2$ are shown against $c$  
from the top to the bottom in blue, red and green lines, respectively, 
for $N=20$. We recall $n=N/L=1$. }
\label{c-cf}
\end{center}
\end{figure}

The estimates of condensate fraction $n_0$ are plotted against 
coupling constant $c$ in the upper panel of Figure \ref{c-cf} 
over a wide range of $c$ such as from $c=10^{-3}$ to $c=10^3$ 
for different values of particle number $N$ 
such as $N=4$, 10, \ldots, 400. 
 For each $N$, condensate fraction $n_0$ becomes 1.0 for small $c$ 
such as $c < 0.01$, while it decreases with respect to $c$ and  
approaches an asymptotic value in the large $c$ region such as 
$c > 100$ or 1000.  
The asymptotic values depend on particle number $N$ 
for $N=4$, 10, \ldots, 400, and they are consistent 
with the numerical estimates of occupation numbers 
for the impenetrable 1D Bose gas (see \Erase{eq.} \Add{Equation} (56) of Ref. \cite{PhysRevA.67.043607}).   
In the lower panel of Figure \ref{c-cf}, 
we plot fractions $n_j$ for $j=0,1$ and 2 against coupling constant 
$c$ from $c=10^{-3}$ to $c=10^3$ with $N=20$. 
The asymptotic values of $n_j$ for large $c$  (i.e. $c=1000$) 
are consistent with the numerical estimates for the impenetrable 1D Bose gas 
(for $n_1$ and $n_2$, see \Erase{eqs.} \Add{Equations} (57) and (58) of Ref. \cite{PhysRevA.67.043607},  
respectively).   

\begin{figure}[t]
\centering
\includegraphics[width=0.6 \textwidth]{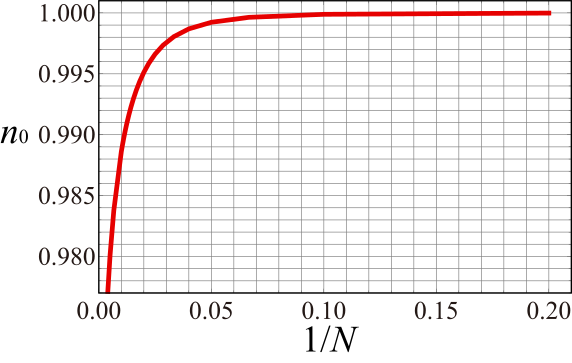} 
\caption{(Color online) Condensate fraction $n_0$ as a function of  
$1/N$ for $c=0.01$. Here $n=N/L=1.0$. 
}
\label{n0-1/N}
\end{figure}

We observe that condensate fraction $n_0$ decreases 
as particle number $N$ increases where density $n=N/L$ is fixed.  
It is the case for $c < 0.1$  in the upper panel of Figure \ref{c-cf}.
Condensate fraction $n_0$ decreases as $N$ increases 
even for small $c$ such as $c=0.01$, as shown in Figure \ref{n0-1/N}. 
Thus, it is necessary for coupling constant $c$ to decrease 
with respect to $N$ so that condensate fraction $n_0$ remains 
constant as $N$ increases with fixed density $n$.

\subsection{Exact finite-size scaling}

\begin{figure}[t]
\begin{center}
\includegraphics[width=0.7\textwidth]{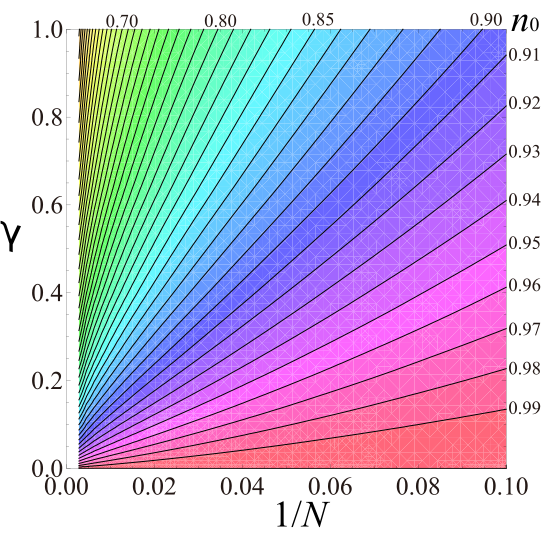}
\caption{(Color online) Contours of condensate fraction 
$n_0$ are plotted for various values of $n_0$ 
in the $\gamma$ versus $1/N$ plane.  
Each contour is approximated by (\ref{eq:scaling}): 
$\gamma$ as a function of $1/N$.   
}
\label{fig:contours}
\end{center}
\end{figure}

We now show the finite-size scaling of condensate fraction $n_0$. 
In Figure \ref{fig:contours} 
each contour line gives the graph of interaction parameter  
$\gamma$ as a function of the inverse of particle number $N$ 
for a fixed value of condensate fraction $n_0$.  
They are plotted for various values of $n_0$ from $n_0=0.6$ to 0.99, 
and are obtained by solving the Bethe-ansatz equations numerically.


For different values of density such as $n=1$, 2 and 5,   
we have plotted contour lines with fixed values of condensate fraction $n_0$ 
in the plane of interaction parameter $\gamma$ versus 
inverse particle number $1/N$. We have observed that the contours with the same condensate fraction $n_0$ but for the different densities coincided with each other in the $\gamma$ versus $1/N$ plane.  
Furthermore, they are well approximated by   
\begin{equation}
\gamma = A/N^{\eta} . \label{eq:scaling}
\end{equation}
Thus, condensate fraction $n_0$ is constant as particle number $N$ becomes very large 
if interaction parameter $\gamma$ is given by  
the power of particle number $N$ as in \Eref{eq:scaling}.

Applying the finite-size scaling arguments, 
we suggest from \Add{Equation} (\ref{eq:scaling}) that condensation fraction 
$n_0$ is given by a scaling function $\phi(\cdot)$ 
of a single variable $\gamma N^{\eta}$: $n_0=\phi(\gamma N^{\eta})$. 
Here we recall the coincidence of contours for the different values of density 
$n$ in \Fref{fig:contours}.  
We thus observe that exponent $\eta$ and amplitude $A$ 
of \Erase{eq.} \Add{Equation} (\ref{eq:scaling}) are determined only by condensate fraction $n_0$ 
and are independent of density $n$.  

\begin{figure}[t]
\begin{center}
\includegraphics[width=0.45\textwidth]{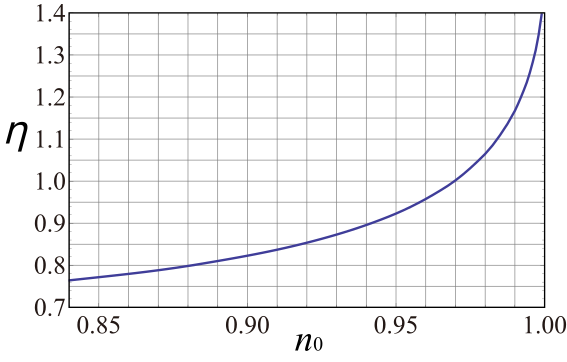}
\includegraphics[width=0.45\textwidth]{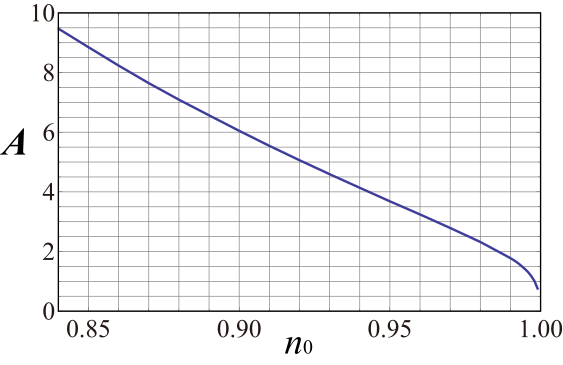}
\caption{(Color online) 
Exponent $\eta$ and amplitude $A$ 
as functions of condensate fraction $n_0$.   
}
\label{fig:eta-A}
\end{center}
\end{figure}

Let us consider amplitude $A$ as a function of $n_0$. 
We denote it by $A=f(n_0)$.   
Then, the scaling function $\phi(\cdot)$ is given by 
the inverse function: $n_0=f^{-1}(A)$.  
In Figure \ref{fig:eta-A}, exponent $\eta$ increases with respect to $n_0$, and 
amplitude $A$ decreases monotonically with respect to $n_0$.

\subsection{Quasi-BEC according to the Onsager-Penrose criterion}

It follows from (\ref{eq:scaling}) that 
BEC does not occur in the 1D Bose gas if we fix 
parameter $\gamma$ and density $n$ as system size $L$ goes to infinity.  
However, if $\gamma$ is small enough 
so that it satisfies \Erase{eq.} \Add{Equation} (\ref{eq:scaling}) 
for a given value of condensate fraction $n_0$, 
the 1D Bose gas shows \Add{the quasi-}BEC from the viewpoint 
of the Penrose and Onsager criterion. 
We suggest that if condensate fraction $n_0$ of a quantum state 
is nonzero and finite for large $N$,  
the mean-field approximation is valid for the quantum state. 
For instance, there exist such quantum states 
that correspond to classical dark-solitons of the GP equation \cite{Sato_2012}, 
if parameter $\gamma$ is small enough so  
that it satisfies \Add{Equation} (\ref{eq:scaling}).

\subsection{Various limiting procedures}

With the scaling behavior \Add{expressed in Equation} (\ref{eq:scaling}) 
we derive various ways of the thermodynamic limit 
such that condensate fraction $n_0$ is constant. 
For instance, we consider 
the case of a finite particle number, $N=N_{\rm f}$. 
Choosing a value of $n_0$, we determine $\gamma$ 
by \Erase{eq.} \Add{Equation} (\ref{eq:scaling}) as $\gamma= A(n_0)/N_{\rm f}^{\eta(n_0)}$.      
Then, the 1D Bose gas with $N=N_{\rm f}$ 
has the same condensate fraction $n_0$ for any large value of $L$ 
if coupling constant $c$ is given by $c=A(n_0) N_{\rm f}^{1- \eta}/L$. 
Let us set $\eta=1$ and $N_{\rm f}=10$, for simplicity. 
We have $n_0=0.97$ in Figure \ref{fig:eta-A},    
and $\gamma = 0.3$ at $1/N=0.1$ in the contour of $n_0=0.97$ 
in Figure \ref{fig:contours}. 
By assuming $n=1$, it corresponds 
to the case of $L=10$ and $c = 0.3$, and we have $A=c L = 3$, 
which is consistent with Figure \ref{fig:eta-A}. 
Therefore, the 1D Bose gas with $N_{\rm f}=10$ has $n_0 = 0.97$ for any large $L$ if $c$ is given by $c=0.3/L$.  
Moreover, we may consider other types of thermodynamic limits. 
When density $n$ is proportional to a power of $L$ as $L^{\alpha}$, 
condensate fraction 
$n_0$ is constant as $L$  goes to infinity if we set 
$c \propto L^{(1-\eta)(1+\alpha)-1}$.

The scaling law \Add {in Equation} (\ref{eq:scaling}) and the estimates of condensate fraction 
in the present paper should be useful for estimating conditions 
in experiments of trapped cold atomic gases in one dimension \cite{ptaevskii2003BEC}. 
For instance, we suggest from Figure \ref{c-cf} that 
BEC may appear in 1D systems with a small number of bosons 
such as $N=20$ or $40$ for $c=1$ or $10$.

\section{Concluding remarks}
In the first part, we have shown that the density profile and the square amplitude evolved in time differently, in particular, for the equal weight case. In the former the notches were filled progressively, while the amplitude of the latter decreased gradually. Furthermore, the Gaussian weights led to the different depths for quantum double dark-solitons \cite{kinjo2022quantum}. This gave the two notches of the quantum double dark-soliton the different speeds, and we observed the scattering of the two notches in the quantum double dark-soliton state exactly. Interestingly, the winding number of the quantum double dark-soliton state has changed when the two notches approach. \Add{Here we recall that it is not necessary for the winding number to be conserved in the time evolution of the quantum system, since it is defined for the corresponding classical system. }

In the second part, 
we  exactly calculated the 
condensate fraction of the 1D Bose gas with repulsive interaction  
by the form factor expansion \Add{for the ground state}. 
We have shown the finite-size scaling behavior 
such that condensate fraction $n_0$ is given by a 
scaling function of interaction parameter $\gamma$ times 
some power of particle number $N$: $n_0=\phi(\gamma N^{\eta})$.  
Consequently, if parameter $\gamma$ decrease as $\gamma=A/N^{\eta}$, 
condensate fraction $n_0$ remains nonzero and constant 
as particle number $N$ becomes very large.   
By modifying the thermodynamic limit, 
the 1D Bose gas shows BEC from the viewpoint of 
the Penrose-Onsager criterion.

\section{Acknowledgements}

The present research is partially supported by Grant-in-Aid 
for Scientific Research No. 21K03398.
K. K. is supported by the Japan Science Technology Agency (CREST Grant Number JPMJCR 19T4).

\section*{References}
\bibliographystyle{iopart-num}
\bibliography{reference}

\end{document}